\documentclass[aps,pra,twocolumn,showpacs,nofootinbib]{revtex4}
\usepackage{amsmath,amssymb,amsbsy,subfigure,times,color}
\usepackage[dvips]{graphicx}
\usepackage{hyperref}

\newcommand{\R}{\mathbb{R}}
\newcommand{\id}{\mathbb{I}}

\newcommand{\be}{\begin{equation}}
\newcommand{\ee}{\end{equation}}
\newcommand{\bea}{\begin{eqnarray}}
\newcommand{\eea}{\end{eqnarray}}

\renewcommand{\det}{{\rm Det}\,}
\newcommand{\gr}[1]{\boldsymbol{#1}}
\newcommand{\ket}[1]{|#1\rangle}

\newcommand{\ketbra}[2]{| #1\rangle\!\langle #2 |}

\newcommand{\N}{{\cal N}}

\newcommand{\sig}{\gr{\sigma}}
\newcommand{\eps}{\gr{\varepsilon}}

\newcommand{\eq}[1]{Eq.~(\ref{#1})}
\newcommand{\ineq}[1]{Ineq.~(\ref{#1})}

\newcommand{\ie}{\emph{i.e.}~}

\begin{document}

\title{Continuous variable entanglement sharing in non-inertial frames}

\date{October 25, 2007}

\author{Gerardo Adesso}
\affiliation{Dipartimento di Fisica ``E. R. Caianiello'',
Universit\`a degli Studi di Salerno, 84081 Baronissi (SA), Italy;
and \\ Grup de F\'isica Te\`orica, Universitat Aut\`onoma de
Barcelona, 08193 Bellaterra (Barcelona), Spain.}
\author{Ivette Fuentes-Schuller}\thanks{Published before under
maiden name Fuentes-Guridi} \affiliation{Instituto de Ciencias
Nucleares, Universidad Nacional Aut\'onoma de M\'exico, A-postal
70-543 04510, Mexico D.F.}
\author{Marie Ericsson} \affiliation{Centre for Quantum Computation, DAMTP, Centre for Mathematical
Sciences, University of Cambridge, Wilberforce Road, Cambridge CB3
0WA, United Kingdom.}


\begin{abstract}We study the distribution of entanglement between modes of a free scalar field
from the perspective of observers in uniform acceleration. We
consider a two-mode squeezed state of the field from an inertial
perspective, and analytically study the degradation of entanglement
due to the Unruh effect, in the cases of either one or both
observers undergoing uniform acceleration. We find that for two
observers undergoing finite acceleration, the entanglement vanishes
between the lowest frequency modes. The loss of entanglement is
precisely explained as a redistribution of the inertial entanglement
into multipartite quantum correlations among accessible and
unaccessible modes from a non-inertial perspective. We show that
classical correlations are also lost from the perspective of two
accelerated observers but conserved if one of the observers remains
inertial.
\end{abstract}

% \tableofcontents
%\title[Continuous variable entanglement sharing in non-inertial frames]
%\shorttitle{Entanglement sharing of Gaussian
%states}
\maketitle

%%%%%%%%%%%%%%%%%%%%%%%%%%%%%%%%%%%%%%%%%%%%%%%%%%%%%%%%%%%%%%%%%%%%%%%%%%%%%%%%%%%%%%%%%%%%%%
\section{Introduction}

In the study of most quantum information tasks such as teleportation
and quantum cryptography, non-relativistic observers share entangled
resources to perform their experiments \cite{nielsen}. Apart from a
few studies \cite{relativistic,alice,dirac,ball,alsing,ahn,zhang},
most works on quantum information assume a world without gravity
where spacetime is flat. But the world is relativistic and any
serious theoretical study must take this into account. It is
therefore of fundamental interest to revise quantum information
protocols in relativistic settings \cite{peresreview}. It has been
shown that relativistic effects on quantum resources are not only
quantitatively important but also induce novel, qualitative features
\cite{alice,dirac,ball,ahn}. For example, it has been shown that the
dynamics of spacetime can generate entanglement \cite{ball}. This,
in principle, would have a consequence in any entanglement-based
protocol performed in curved spacetime. Relativistic effects have
also been found to be relevant in a flat spacetime, where the
entanglement described by observers in uniform acceleration is
observer-dependent since it is degraded by the Unruh effect
\cite{alice,dirac,ahn}. In the infinite acceleration limit, the
entanglement vanishes for bosons \cite{alice,ahn} and reaches a
non-vanishing minimum for fermions \cite{dirac}. This degradation on
entanglement results in the loss of fidelity of teleportation
protocols which involve observers in uniform acceleration
\cite{alsing}.

Understanding entanglement in a relativistic framework is not only
of interest to quantum information. Entanglement plays an important
role in black hole entropy \cite{bombelli} and in the apparent loss
of information in black holes \cite{loss}, one of the most
challenging problems in theoretical physics at the moment
\cite{information}. Understanding the entanglement between modes of
a field close to the horizon of a black hole might help to
understand some of the key questions in black hole thermodynamics
and their relation to information.

In this paper we interpret the loss of bipartite entanglement
between two modes of a scalar field from a non-inertial perspective,
as an effect of entanglement redistribution. Precisely, we consider
the entanglement between two field modes, each described from the
perspective of a different observer. Suppose that the two modes are
entangled to a given degree from the perspective of two inertial
observers. The state will appear less entangled if either one or
both the observers move with uniform acceleration \cite{alice}. This
is because each mode described by an inertial observer corresponds
to two entangled modes from the perspective of a non-inertial
observer \cite{unruh}. Consequently, a two-mode entangled state
described from the inertial perspective, corresponds to a three-mode
state when described from the perspective of one inertial observer
and one in uniform acceleration, and to a four-mode state if both
observers are accelerated. Physical observers moving with uniform
acceleration have access only to one of the non-inertial modes.
Thus, when describing the state (which involves tracing over the
unaccessible modes) the observers find that some of the correlations
are lost.

This phenomenon, stemming from the Unruh effect \cite{unruh}, was
first studied from the quantum information perspective for bosonic
scalar fields \cite{alice} (considering one inertial observer and
the other one undergoing uniform acceleration) and later for
fermionic Dirac fields \cite{dirac}. Although entanglement of
particle number states is in both cases degraded as a function of
the acceleration, there are important differences in the results.
For example, in the infinite acceleration limit, the entanglement
reaches a non-vanishing minimum value for fermions, while it
completely disappears in the bosonic case. For photon helicity
entangled states, instead, the correlations are not degraded at all
\cite{zhang}. The loss of entanglement was explained in the
fermionic case in the light of the entanglement sharing framework as
an effect of the redistribution of entanglement among all,
accessible and unaccessible, modes. Although the loss of
entanglement was first studied for scalar fields (considering a
 state from an inertial perspective which is maximally
entangled in a two-qubit space, $\ket{\psi} \sim
\ket{00}+\ket{11}$), entanglement sharing was not analyzed in that
instance, due to the difficulty of computing entanglement in such a
hybrid qubit--continuous-variable system. Fortunately, the theory of
continuous variable entanglement has been in recent times developed,
allowing for the exact, quantitative study of bipartite entanglement
and its distribution in the special class of Gaussian states
\cite{adebook}, which includes, among others, squeezed, coherent and
thermal states of harmonic oscillators.

Here, we consider a free scalar field which is, from an inertial
perspective, in a two-mode squeezed state. This choice of the state
is motivated by different observations. First, the two-mode squeezed
state is the paradigmatic entangled state of a continuous variable
system, approximating to an arbitrarily good extent the
Einstein-Podolski-Rosen (EPR) pair \cite{epr}. Second, the state can
be produced in the lab and exploited for any current realization of
bipartite quantum information with continuous variables
\cite{brareview}. Third, it belongs to the class of Gaussian states,
which admit an exact description of their classical and quantum
correlations. Since the Unruh transformations \cite{unruh} are
Gaussian themselves (\ie they preserve the Gaussian character of the
state), it is possible to characterize analytically the
redistribution of correlations due to relativistic effects. Finally,
the two-mode squeezed state has a special role in quantum field
theory. It is possible to define particle states (necessary in any
entanglement discussion) when the spacetime has at least two
asymptotically flat regions \cite{birelli,ball}. In this case,
particle states commonly correspond to multi-mode squeezed states in
which several field modes are in a pair-wise squeezed entangled
state. The state we consider in our entanglement discussion is the
simplest multi-mode squeezed state in which only two modes are
entangled.

A first investigation on the degradation of entanglement in a
two-mode squeezed state due to the Unruh effect has been recently
reported \cite{ahn}. The entanglement degradation (quantified by the
logarithmic negativity \cite{vidwer}) was analyzed when one of the
observers is accelerated and found to decrease more drastically when
the entanglement described from the inertial perspective is stronger
and to vanish in the infinite acceleration limit.

We perform an extensive study of both quantum (entanglement) and
classical correlations of the two-mode squeezed state  from a
non-inertial perspective. Our work aims at a conclusive
understanding and characterization of the relativistic effects on
continuous variable correlations described by observers in uniform
acceleration. Therefore, we evaluate not only the bipartite
entanglement as degraded by the Unruh thermalization, but
remarkably, the multipartite entanglement which arises among all
Rindler modes. Our analysis is possible thanks to recent analytical
results on entanglement sharing and the quantification of
multipartite entanglement in Gaussian states. This analysis relays
on the {\em contangle}, which is a computable measure of
entanglement \cite{contangle}. The contangle for mixed states is not
fully equivalent to the negativity. Therefore, in the case of a
single accelerated observer, our results will evidence significant
differences with the results presented in Ref.~\cite{ahn}. The main
novel result we find in this case, is that in the infinite
acceleration limit, all the bipartite entanglement described by
inertial observers is exactly redistributed into genuine tripartite
correlations between the modes described by one inertial and two
non-inertial observers (one real and one fictitious, or virtual), as
a consequence of the monogamy constraints on entanglement
distribution \cite{contangle,hiroshima,pisa}. We also analyze total
correlations, finding that the classical correlations are invariant
under acceleration when one observer is accelerated.

Furthermore, we present an original analysis of the Unruh effect on
continuous variable entanglement when both observers undergo uniform
acceleration. This analysis yields a series of significant new
results. First, the bipartite entanglement described by non-inertial
observers may vanish completely at finite acceleration even when the
state contains an infinite amount of entanglement from the point of
view of inertial observers. Second, the acceleration induces a
redistribution of entanglement, such that the modes described from a
non-inertial perspective are correlated via a genuine four-partite
entanglement. This entanglement increases unboundedly with the
acceleration, easily surpassing the original inertial bipartite
entanglement. Third, classical correlations are also degraded as
function of the acceleration. The degradation is of at most one unit
with respect to the case of a single non-inertial observer.
Moreover, we study the dependence of the bipartite entanglement on
the frequency of the modes described by the non-inertial observers,
finding that with increasing acceleration the range of entangled
frequencies gets narrower and narrower, becoming empty in the limit
of infinite acceleration.

Our results are on one hand an interesting application of the
continuous variable quantum information techniques (commonly
confined to quantum optics or light--matter interfaces) to a
relativistic setting, and on the other hand, provide a deeper
understanding of the characterization of the inherent relativistic
effects on the distribution of information. This may lead to a
better understanding of the behavior of information in presence of a
black hole \cite{letter}.

The paper is organized as follows. In Section \ref{secGauss} we
introduce the basic tools of quantum information with Gaussian
states of continuous variable systems and we discuss the mechanism
of entanglement sharing. In Section \ref{SecUnruh}  we describe the
Unruh effect and its consequences on the entanglement  between two
field modes. In Section \ref{secOne}, we study distributed
entanglement between modes of a free scalar field when one observer
is accelerated. The case when both observers are accelerated,
resulting in a four-partite entangled state, is studied in Section
\ref{secTwo}. Both Sections \ref{secOne} and \ref{secTwo} include an
analysis of the dependence of classical correlations under
acceleration of the observers. Finally, in Section \ref{SecConcl} we
draw our concluding remarks and compare our results to those
obtained in the case of Dirac fields \cite{dirac}.

\section{Preliminary toolbox}\label{secPrelim}

\subsection{Gaussian states and Gaussian entanglement
measures}\label{secGauss}

Entanglement in continuous variable (CV) systems is encoded in the
form of Einstein-Podolski-Rosen (EPR) correlations \cite{epr}. Let
us consider the quadratures of a two-mode radiation field,  where
mode $k=i,j$ is described by the ladder operators $\hat {a}_k
,\,\hat {a}_k ^\dag $ satisfying the bosonic commutation relation
$[\hat {a}_k ,\,\hat {a}_k ^\dag ]=1$. An arbitrarily increasing
degree of entanglement can be encoded in a two-mode squeezed state
$\ket{\psi^{sq}}_{i,j}=U_{i,j}(r)
\left(\ket{0}_i\!\otimes\ket{0}_j\right)$ with increasing squeezing
factor $r \in \R$, where  the (phase-free) two-mode squeezing
operator is given by
\begin{equation}\label{tmsU}
U_{i,j}(r) = \exp \left[\frac{r}{2} (\hat {a}_i^\dag \hat {a}_j^\dag
-\hat {a}_i \hat {a}_j ) \right]\,,
\end{equation} and $\left| 0
\right\rangle _k $ denotes the vacuum state in the Fock space of
mode $k$. In the limit of infinite squeezing ($r\to \infty )$, the
state approaches the ideal EPR state \cite{epr} which is,
simultaneously, eigenstate of total momentum and relative position
of the two subsystems. Therefore, the state contains infinite
entanglement. The EPR state is unnormalizable and unphysical.
 The two-mode squeezed state is an arbitrarily good approximation of
it with increasing squeezing, and therefore represents a key
resource for practical implementations of CV quantum information
protocols \cite{brareview}. Mathematically, squeezed states belong
to the class of {\em Gaussian states} of CV systems, {\em
i.e.}~states with Gaussian characteristic functions and
quasi-probability distributions, whose structural and informational
properties have been intensively studied in recent times
\cite{adebook}.

\subsubsection{Covariance matrix formalism}

In view of the subsequent analysis, it is sufficient to recall that
Gaussian states of $N$ modes are completely described in phase space
(up to local unitaries) by the real, symmetric covariance matrix
(CM) $\gr{\sigma}$, whose entries are
$\sigma_{ij}=1/2\langle\{\hat{X}_i,\hat{X}_j\}\rangle
-\langle\hat{X}_i\rangle\langle\hat{X}_j\rangle$. Here
$\hat{X}=\{\hat x_1,\hat p_1,\ldots,\hat x_N,\hat p_N\}$ is the
vector of the field quadrature operators, whose canonical
commutation relations can be expressed in matrix form: $[\hat
X_{i},\hat X_j]=2i\Omega_{ij}$, with the symplectic form
$\Omega=\oplus_{i=1}^{n}\omega$ and $\omega=\delta_{ij-1}-
\delta_{ij+1},\, i,j=1,2$.  The CM $\gr{\sigma}$ must fulfill the
Robertson-Schr\"odinger uncertainty relation \cite{simon87}
\begin{equation}\label{bonfide}
\gr{\sigma}+i\Omega \geq 0\,,
\end{equation}
to describe a physical state. Throughout the paper, $\gr{\sigma}$
will be used indifferently to indicate the CM of a Gaussian state or
the state itself.

Unitary Gaussian operations $U$ amount, in phase space, to
symplectic transformations $S$ (which preserve the symplectic form,
$\Omega=S^T \Omega S$) acting ``by congruence'' on the CM ({\ie}so
that $\sig\mapsto S \sig S^T$). For instance, the two-mode squeezing
operator \eq{tmsU} corresponds to the symplectic transformation
\begin{equation}\label{tmsS}
S_{i,j}(r)=\left(\begin{array}{cccc}
\cosh r&0&\sinh r&0\\
0&\cosh r&0&-\sinh r\\
\sinh r&0&\cosh r&0\\
0&-\sinh r&0&\cosh r
\end{array}\right)\, ,
\end{equation}
where the matrix is understood to act on the couple of modes $i$ and
$j$. A two-mode squeezed state with squeezing degree $r$
\cite{walls} will be thus described by a CM
\begin{eqnarray}\label{tms}
\sig^{sq}_{i,j}(r)&=&S_{i,j}(r) \id_{4} S_{i,j}^T(r)
\\ &=&\left(\begin{array}{cccc}
\cosh2r&0&\sinh2r&0\\
0&\cosh2r&0&-\sinh2r\\
\sinh2r&0&\cosh2r&0\\
0&-\sinh2r&0&\cosh2r
\end{array}\right)\! , \nonumber
\end{eqnarray}
where we have used that the CM of a $N$-mode vacuum is the $2N\times
2N$ identity matrix $\id_{2N}$.

\subsubsection{Qualifying and quantifying entanglement}

For what concerns characterizing bipartite entanglement, the
positive partial transpose (PPT) criterion states that a Gaussian CM
$\sig$ is separable (with respect to a $1 \times N$ bipartition) if
and only if the partially transposed CM $\tilde{\sig}$ satisfies the
uncertainty principle given by \eq{bonfide} \cite{simon,werwolf}.
The suffix ``$\sim$'' denotes the partial transposition, implemented
by reversing time in the subspace of only one subsystem of a
bipartite composite CV system \cite{simon}. An ensuing computable
measure of CV entanglement is the \emph{logarithmic negativity}
\cite{vidwer} $E_{\N}\equiv \log\|\tilde{\varrho}\|_{1}$, where $\|
\cdot \|_1$ denotes the trace norm. This measure is an upper bound
to the {\em distillable entanglement} of the state $\varrho$. The
logarithmic negativity is used in Ref.~\cite{ahn} to quantify the
degradation of two-mode Gaussian entanglement due to one accelerated
observer.

We employ a different measure of bipartite entanglement: the {\em
contangle} \cite{contangle}, which is an entanglement monotone under
Gaussian local operations and classical communication (GLOCC), that
belongs to the family of `Gaussian entanglement measures'
\cite{ordering}. The principal motivation for this choice is that
our main focus is to study the effects of the Unruh thermalization
mechanism on the distribution of entanglement among field modes
described from a non-inertial perspective. In this setting, the
contangle is {\em the} measure enabling a mathematical treatment of
distributed CV entanglement as emerging from the fundamental
monogamy constraints \cite{contangle,hiroshima,pisa}. The contangle
$\tau$ is defined for pure states as the square of the logarithmic
negativity and it is extended to mixed states via the Gaussian
convex roof \cite{ordering,geof}, that is as the minimum of the
average pure-state entanglement over all decompositions of the mixed
state in ensembles of pure Gaussian states. If $\sig_{i\vert j} $ is
the CM of a (generally mixed) bipartite Gaussian state where
subsystem $i$ comprises one mode only, then the contangle $\tau$ can
be computed as \cite{contangle}
\begin{equation}
\label{tau} \tau (\sig_{i\vert j} )\equiv \tau (\sig_{i\vert
j}^{opt} )=g[m_{i\vert j}^2 ],\;\;\;g[x]={\rm arcsinh}^2[\sqrt {x-1}
],
\end{equation}
where $\sig_{i\vert j}^{opt} $ corresponds to a pure Gaussian state,
and $m_{i\vert j} \equiv m(\sig_{i\vert j}^{opt} )=\sqrt {\det
\sig_i^{opt} } =\sqrt {\det \sig_j^{opt} } $, with
$\sig_{i(j)}^{opt} $ the reduced CM of subsystem $i (j)$, obtained
tracing over the degrees of freedom of subsystem $j$ ($i)$. The CM
$\sig_{i\vert j}^{opt} $ denotes the pure bipartite Gaussian state
which minimizes $m(\sig_{i\vert j}^p )$ among all pure-state CMs
$\sig_{i\vert j}^p $ such that $\sig_{i\vert j}^p \le \sig_{i\vert
j} $. If $\sig_{i\vert j} $ is a pure state, then $\sig_{i\vert
j}^{opt} =\sig_{i\vert j} $, while for a mixed Gaussian state
\eq{tau} is mathematically equivalent to constructing the Gaussian
convex roof. For a separable state, $m(\sig_{i\vert j}^{opt} )=1$
and the entanglement vanishes. The contangle $\tau$ is completely
equivalent to the Gaussian entanglement of formation \cite{geof},
which quantifies the cost of creating a given mixed, entangled
Gaussian state out of an ensemble of pure, entangled Gaussian
states. Notice also that in general the Gaussian entanglement
measures are inequivalent to the negativities, in that they may
induce opposite ordering on the set of entangled, nonsymmetric
two-mode Gaussian states \cite{ordering}: this will be explicitly
unfolded in the following analysis.

\subsubsection{Entropy and mutual information}

In a bipartite setting, another important correlation measure is the
so-called {\em mutual information}  quantifying the total (classical
and quantum) correlations between two parties. The mutual
information of a state $\varrho_{A|B}$ of a bipartite system is
defined as
\begin{equation}\label{mi}
I(\varrho_{A|B}) = S_V(\varrho_A) + S_V(\varrho_B) -
S_V(\varrho_{A|B})\,,
\end{equation}
where $\varrho_A$ ($\varrho_B$) is the reduced state of subsystem
$A$ ($B$) and $S_V$ denotes the Von Neumann entropy, defined for a
quantum state $\varrho$ as $S_V(\varrho) = - {\rm Tr} \varrho \log
\varrho$. If $\varrho_{A|B}$ is a pure quantum state
[$S_V(\varrho_{A|B})=0$], the Von Neumann entropy of its reduced
states $S_V(\varrho_A)=S_V(\varrho_B)$ quantifies the entanglement
between the two parties \cite{noteentro}. Being
$I(\varrho_{A|B})=2S_V(\varrho_{A})=2S_V(\varrho_{B})$ in this case,
one says that the pure state also contains some classical
correlations, equal in content to the quantum part,
$S_V(\varrho_A)=S_V(\varrho_B)$. In mixed states, the distinction
between classical and quantum correlations cannot be considered an
accomplished task yet \cite{henvedral01}.

For an arbitrary bipartite (pure or mixed) Gaussian state, the Von
Neumann entropy and hence the mutual information can be easily
computed in terms of the symplectic spectra of the CM of the global
state, and of the reduced CMs of both subsystems. In the case of a
two-mode state with global CM $\sig_{A|B}$, the mutual information
yields \cite{HolevoWerner,Serafini}
\begin{equation}\label{migau}
I(\sig_{A|B}) = f(\sqrt{\det\sig_A})+f(\sqrt{\det\sig_B}) -
f(\eta^-_{A|B}) - f(\eta^+_{A|B})\,,
\end{equation}
where \begin{equation}\label{entfunc} f(x) \equiv
\frac{x+1}{2}\log\left(\frac{x+1}{2}\right)-
\frac{x-1}{2}\log\left(\frac{x-1}{2}\right) \, ,
\end{equation}
and $\{\eta^-_{A|B},\,\eta^+_{A|B}\}$ are the symplectic eigenvalues
of $\sig_{A|B}$ (\ie the orthogonal eigenvalues of the matrix $|i
\Omega \sig_{A|B}|$).

\subsubsection{Distributed quantum correlations and multipartite
entanglement}

Quantifying entanglement in multipartite systems is generally very
involved. A way to determine the existence of multipartite
correlations in a state is by exploring the entanglement distributed
between multipartite systems. Unlike classical correlations,
entanglement is {\em monogamous}, meaning that it cannot be freely
shared among multiple subsystems of a composite quantum system
\cite{pisa}. This fundamental constraint on entanglement sharing has
been mathematically demonstrated, so far, for arbitrary systems of
qubits within the discrete-variable scenario \cite{ckw,osborne}, for
a special case of two qubits and an infinite-dimensional system
\cite{trace}, and for all $N$-mode Gaussian states within the CV
scenario \cite{contangle,hiroshima}.

In the general case of a state distributed among $N$ parties (each
owning a single qubit, or a single mode, respectively), the monogamy
constraint takes the form of the Coffman-Kundu-Wootters inequality
\cite{ckw},
\begin{equation}
\label{ckwine} E_{S_i \vert (S_1 \ldots S_{i-1} S_{i+1} \ldots S_N
)} \ge \sum\limits_{j\ne i}^N {E_{S_i \vert S_j } },
\end{equation}
where the global system is multipartitioned in subsystems $S_k$
($k=1,{\ldots},N$), each owned by a respective party, and $E$ is a
proper measure of bipartite entanglement. The left-hand side of
inequality (\ref{ckwine}) quantifies the bipartite entanglement
between a probe subsystem $S_i $ and the remaining subsystems taken
as a whole. The right-hand side quantifies the total bipartite
entanglement between $S_i$ and each one of the other subsystems
$S_{j\ne i}$ in the respective reduced states. The non-negative
difference between these two entanglements, minimized over all
choices of the probe subsystem, is referred to as the
\textit{residual multipartite entanglement}. It quantifies the
purely quantum correlations that are not encoded in pairwise form,
so it includes all manifestations of genuine $K$-partite
entanglement, involving $K$ subsystems at a time, with $2<K\le N$.
In the simplest nontrivial instance of $N=3$, the residual
entanglement has the meaning of the genuine tripartite entanglement
shared by the three subsystems \cite{ckw}. Such a quantity has been
proven to be a tripartite entanglement monotone for pure three-mode
Gaussian states, when bipartite entanglement is quantified by the
contangle \cite{contangle}.

\subsection{Entanglement in non-inertial frames: the Unruh effect}\label{SecUnruh}

To study entanglement from the point of view of parties in uniform
acceleration it is necessary to consider that field quantization in
different coordinates is inequivalent. While an inertial observer
concludes that the field is in the vacuum state, from the
perspective of an observer in uniform acceleration the field is
described as a thermal distribution of particles proportional to
his/her acceleration. This is known as the Unruh effect \cite{unruh}
and it has important consequences on the entanglement between
(bosonic and/or fermionic) field modes and its distribution
properties \cite{alice,dirac}. We will study such consequences in
the case of a bosonic field in a state which corresponds to a
two-mode squeezed state from an inertial perspective (see also
\cite{ahn}). Let us first discuss how the Unruh effect arises.

Consider an observer moving in the $(t,z)$ plane ($c=1$) with
uniform acceleration $\aleph$. Rindler coordinates $(\tau,\zeta)$
are appropriate for describing the viewpoint of an uniformly
accelerated observer. Two different sets of Rindler coordinates,
which differ from each other by an overall change in sign, are
necessary for covering Minkowski space,
\begin{eqnarray*}\label{Rindler_coords}
\aleph t &=&  e^{\aleph\zeta}\sinh(\aleph\tau), \quad \aleph z =
e^{\aleph\zeta}\cosh(\aleph\tau),\\
 &&\\  \aleph t &=&
-e^{\aleph\zeta}\sinh(\aleph\tau), \quad \aleph z =
-e^{\aleph\zeta}\cosh(\aleph\tau).
\end{eqnarray*}
These sets of coordinates define two Rindler regions (respectively
$I$ and $II$) that are causally disconnected from each other. A
particle undergoing eternal uniform acceleration remains constrained
to either Rindler region $I$ or $II$ and has no access to the
opposite region.

Now consider a free quantum scalar field in a flat background. The
quantization of a scalar field in the Minkowski coordinates is not
equivalent to its quantization in Rindler coordinates. However, the
vacuum state of a given field mode described by an inertial observer
can be expressed as a two-mode squeezed state \cite{walls} from the
Rindler perspective \cite{unruh,birelli}
\begin{equation} \label{eq:vacuum}
\left| 0\right\rangle_{\rho_M} = \frac{1}{\cosh r}
\sum_{n=0}^{\infty }\tanh ^{n}r\,\left| n\right\rangle
_{\rho_I}\!\left| n\right\rangle _{\rho_{II}}=U(r)\left|
n\right\rangle _{\rho_I}\!\left| n\right\rangle _{\rho_{II}},
\end{equation}
where
\begin{equation}\label{accparam}
\cosh r
=\left(1-e^{-\frac{2\pi|\omega_\rho|}{\aleph}}\right)^{-\frac12}\,,
\end{equation}
and $U(r)$ is the two-mode squeezing operator introduced in
Eq.~(\ref{tmsU}). Each Minkowski mode of frequency $|\omega_\rho|$
has a Rindler mode expansion given by Eq.~(\ref{eq:vacuum}). The
relation between higher energy states can be found using
Eq.~(\ref{eq:vacuum}) and the Bogoliubov transformation between the
creation and annihilation operators, $$\hat a_{\rho}=\cosh{r}\hat
b_{\rho_I}-\sinh{r}\hat b^{\dagger}_{\rho_{II}},$$ where $\hat
a_{\rho}$ is the annihilation operator in Minkowski space for mode
$\rho$ and $\hat b_{\rho_{I}}$ and $\hat b_{\rho_{II}}$ are the
annihilation operators for the same mode in the two Rindler regions
\cite{unruh}. A Rindler observer moving in region $I$ needs to trace
over the modes in region $II$ since he has no access to the
information in this causally disconnected region. Therefore, while a
Minkowski observer concludes that the field mode $\rho$ is in the
vacuum $|0\rangle_{\rho_M}$, the state from the perspective of an
observer in uniform acceleration $\aleph$, constrained to region
$I$, is
\begin{eqnarray}
\ketbra{0}{0}_{\rho_M}\rightarrow
{1\over\cosh^2r}\sum_{n=0}^\infty\tanh^{2n}r\ketbra{n}{n}_{\rho_I}\,,
\end{eqnarray}
which is a thermal state with temperature $T=\frac{\aleph}{2\pi
k_B}$ where $k_B$ is Boltzmann's constant.

\section{Distributed Gaussian entanglement due to one accelerated
observer}\label{secOne} From the perspective of inertial observers,
we consider a scalar field which is in a two-mode squeezed state
with mode frequencies $\alpha$ and $\rho$ and squeezing parameter
$s$, as in \cite{ahn}. This state, which is the simplest multi-mode
squeezed state (of relevance in quantum field theory
\cite{birelli}), allows for the exact quantification of entanglement
in all partitions of the system from the inertial and non-inertial
perspective. We can define the two-mode squeezed state, described
from an inertial perspective, via its CM [see \eq{tms}]
\begin{equation}\label{inAR}
\sig^{P}_{AR}(s)=S_{\alpha_M,\rho_M}(s)\id_{4}S_{\alpha_M,\rho_M}^T(s)\,,
\end{equation}
where $\id_{4}$ is the CM of the vacuum
$\ket{0}_{\alpha_M}\!\otimes\!\ket{0}_{\rho_R}$.

\begin{figure}[t!]
\centering{\includegraphics[width=7cm]{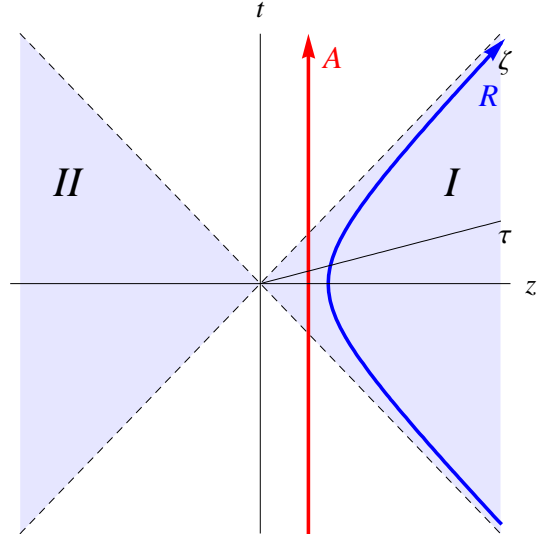}%
\caption{\label{mondouno}(color online) Sketch of the world lines
for the inertial observer Alice and the accelerated observer Rob.
The set $(z,t)$ denotes Minkowski coordinates, while the set
$(\zeta, \tau)$ denotes Rindler coordinates. The causally
disconnected Rindler regions $I$ and $II$ are evidenced.}}
\end{figure}

If an observer (Rob) undergoes uniform acceleration $\aleph_R$, the
state corresponding to the mode $\rho$ \cite{notemodefreq} must be
described in Rindler coordinates (see Fig.~\ref{mondouno}), so that
the Minkowski vacuum is given by
$\ket{0}_{\rho_M}=U_{\rho_I,\rho_{II}}(r)
\left(\ket{0}_{\rho_I}\!\otimes\!\ket{0}_{\rho_{II}}\right)$, with
$U(r)$ given by \eq{tmsU}. Namely, due to the fact that Rob is in
uniform acceleration, the description of the state from his
perspective must include a further two-mode squeezing
transformation, with squeezing $r$ proportional to Rob's
acceleration $\aleph_R$ via \eq{accparam}. As a consequence of this
transformation, the original two-mode entanglement in the state
\eq{inAR} described by Alice (always inertial) and Rob from an
inertial perspective, becomes distributed among the modes described
by Alice, the accelerated Rob moving in Rindler region $I$, and a
virtual anti-Rob ($\bar R$) theoretically able to describe the mode
$\rho_{II}$ in the complimentary Rindler region $II$. Our aim is to
investigate the distribution of entanglement induced by the purely
relativistic effect of Rob's acceleration. It is clear that the
three-mode state described by Alice, Rob and anti-Rob is obtained
from the vacuum by the application of Gaussian unitary operations
only, therefore, it is a pure Gaussian state. Its CM, according to
the above description, is (see also \cite{ahn})
\begin{eqnarray}\label{in3}
\sig_{AR \bar R}(r,s)&=&[\id_{\alpha_M} \oplus
S_{\rho_I,\rho_{II}}(r)]\cdot[S_{\alpha_M,\rho_I}(s) \oplus
\id_{\rho_{II}}]\nonumber \\
&\ \ \cdot&\id_{6}\cdot[S_{\alpha_M,\rho_I}^T(s) \oplus
\id_{\rho_{II}}][\id_{\alpha_M} \oplus S_{\rho_I,\rho_{II}}^T(r)]\,,
\nonumber \\
\end{eqnarray}
where the symplectic transformations $S$ are given by \eq{tmsS}, and
 $\id_{6}$ is the CM of the vacuum
$\ket{0}_{\alpha_M}\!\otimes\!\ket{0}_{\rho_I}\!\otimes\!\ket{0}_{\rho_{II}}$.
Explicitly,
\begin{equation}\label{sig3}
\sig_{AR \bar
R}={\left(%
 \begin{array}{ccc}
  \sig_A& \eps_{AR} &  \eps_{A \bar R} \\
  \eps^T_{AR} & \sig_R & \eps_{R \bar R}\\
  \eps^T_{A \bar R}  & \eps^T_{R \bar R} & \sig_{\bar R} \\
\end{array}%
\right)}\,,
\end{equation}
where:
\begin{eqnarray*}
&& \sig_A = \cosh (2 s) \id_2\,, \\ && \sig_R = [\cosh (2 s) \cosh
^2(r) + \sinh ^2(r)] \id_2\,, \\ && \sig_{\bar R} = [\cosh ^2(r) +
\cosh (2 s) \sinh ^2(r)] \id_2\,, \\ &&\eps_{AR}=[\cosh ^2(r) +
\cosh (2 s) \sinh ^2(r)] Z_2\,, \\ &&\eps_{A \bar R} = [\sinh (r)
\sinh (2 s)] \id_2\,, \\ && \eps_{R \bar R} = [\cosh ^2(s) \sinh (2
r)] Z_2\,,
\end{eqnarray*}
 with $Z_2={{1\ \ \ 0}\choose {0 \ -1}}$.

As pointed out in Ref.~\cite{alice}, the regime of very high
acceleration ($r \gg 0$) can be interpreted as Alice and Rob moving
close to the horizon of a Schwarzschild black hole. While Alice
falls into the black hole, Rob barely escapes the fall by
accelerating away from it with uniform acceleration parameter $r$.

\subsection{Bipartite entanglement}
The contangle $\tau(\sig^P_{A|R})$, quantifying the bipartite
entanglement described by two inertial observers, is equal to
$4s^2$, as can be straightforwardly found by inserting
$m^P_{A|R}=\cosh(2s)$ in \eq{tau}.

Let us now compute the bipartite entanglement in the various $1
\times 1$ and $1 \times 2$ partitions of the state $\sig_{AR \bar
R}$. The $1 \times 2$ contangles are immediately obtained from the
determinants of the reduced single-mode states of the globally pure
state $\sig_{AR \bar R}$, \eq{sig3}, yielding \cite{notem}
\begin{eqnarray}\label{m3_12}
  m_{A|(R \bar R)} &=& \sqrt{\det\sig_A} = \cosh(2s)\,, \\
  m_{R|(A \bar R)} &=& \sqrt{\det\sig_R} = \cosh (2 s) \cosh
^2(r) + \sinh ^2(r)\,, \nonumber \\
  m_{\bar R|(AR)} &=& \sqrt{\det\sig_{\bar R}} = \cosh ^2(r) +
\cosh (2 s) \sinh ^2(r)\nonumber\,.
\end{eqnarray}
For any nonzero value of the two squeezing parameters $s$ and $r$
(\ie entanglement from the point of view of inertial observers and
Rob's acceleration, respectively), each single party is in an
entangled state with the block of the remaining two parties, with
respect to all possible global splitting of the modes. This
classifies the state $\sig_{AR \bar R}$ as {\em fully inseparable}
\cite{barbarella}: it contains therefore genuine tripartite
entanglement, which will be precisely quantified in the next
subsection. Notice also that $m_{A|(R \bar R)}=m^P_{A|R}$, \ie all
the inertial entanglement is distributed, from a non-inertial
perspective, between modes described by Alice and the group \{Rob,
anti-Rob\}, as expected since the coordinate transformation
$S_{\rho_I,\rho_{II}}(r)$ is a local unitary operation with respect
to the considered bipartition, which preserves entanglement by
definition. In the following, we will always assume $s \neq 0$ to
rule out trivial circumstances.

\begin{figure}[t!]
\centering{\includegraphics[width=8.5cm]{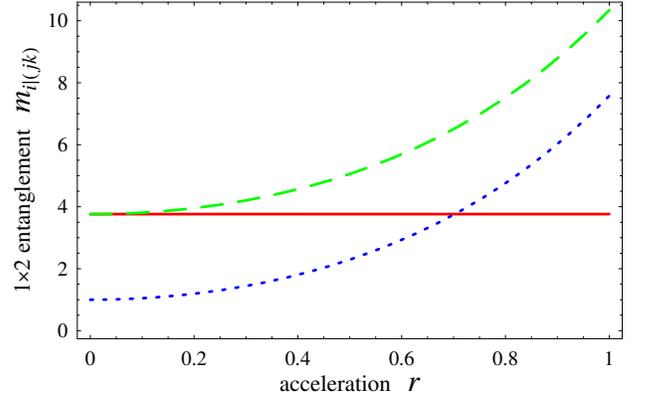}%
\caption{\label{mmshow}(color online) Plot, as a function of the
acceleration parameter $r$, of the bipartite entanglement between
the mode described by one observer and the group of modes described
by the other two, as expressed by the single-mode determinants
$m_{i(jk)}$ defined in \eq{m3_12}. The inertial entanglement is kept
fixed at $s=1$. The solid red line represents $m_{A|(R \bar R)}$,
the dashed green line corresponds to $m_{R|(A \bar R)}$, while the
dotted blue line depicts $m_{\bar R|(AR)}$.}}
\end{figure}

Interestingly, as already pointed out in Ref.~\cite{ahn}, the mode
described by Alice is not directly entangled with the mode described
by anti-Rob, because the reduced state $\sig_{A|\bar R}$ is
separable by inspection, being $\det \eps_{A \bar R} \ge 0$.
Actually, we can further explore this point by noticing that the
mode described by anti-Rob has the {\em minimum} possible bipartite
entanglement with the group of modes described by Alice and Rob.
This follows by recalling that, in any pure three-mode Gaussian
state $\sig_{123}$, the local single-mode determinants have to
satisfy a triangle inequality \cite{3modi}
\begin{equation}\label{triangle}
\left|m_1-m_2\right|+1 \le m_3 \le m_1+m_2-1\,,\end{equation} with
$m_i \equiv \sqrt{\det{\sig_i}}$. In our case, identifying mode 1
with Alice, mode 2 with Rob, and mode 3 with anti-Rob, \eq{m3_12}
shows that the state $\sig_{AR \bar R}$ saturates the leftmost side
of the triangle inequality~(\ref{triangle}), $$m_{\bar
R|(AR)}=m_{R|(A \bar R)}-m_{A|(R \bar R)}+1\,.$$ In other words, the
mixedness of anti-Rob's mode, which is directly related to its
entanglement with the other two modes, is the smallest possible one.
The values of the entanglement parameters $m_{i|(jk)}$ from
\eq{m3_12} are plotted in Fig.~\ref{mmshow} as a function of the
acceleration $r$, for a fixed degree of initial squeezing $s$.

On the other hand, the PPT criterion states that the reduced
two-mode states $\sig_{A|R}$ and $\sig_{R|\bar R}$ are both
entangled. To compute the contangle in those partitions, we first
observe that all the two-mode reductions of $\sig_{AR \bar R}$
belong to the special class of GMEMMS \cite{extremal}, mixed
Gaussian states of maximal entanglement at given marginal
mixednesses. This is a curious coincidence because, when considering
entanglement of Dirac fields from a non-inertial perspective
\cite{dirac}, and describing the effective three-qubit states
described by the three observers, also in that case all two-qubit
reduced states belong to the corresponding family of MEMMS
\cite{memms}, mixed two-qubit states of maximal entanglement at
fixed marginal mixednesses. Back to the CV case, this observation is
useful as we know that for two-mode GMEMMS the Gaussian entanglement
measures, including the contangle, are computable in closed form
\cite{ordering},
\begin{eqnarray}
m_{A|R} &=& \frac{2 \sinh ^2(r) + (\cosh (2 r) +
              3) \cosh (2 s)}{2 \cosh (2 s) \sinh ^2(r) + \cosh (2 r) +
              3}\,, \label{m3_ar} \\
              m_{R|\bar R} &=& \cosh(2r) \label{m3_rr}\,.
\end{eqnarray}

\begin{figure*}[t!]
\centering{\includegraphics[width=13.5cm]{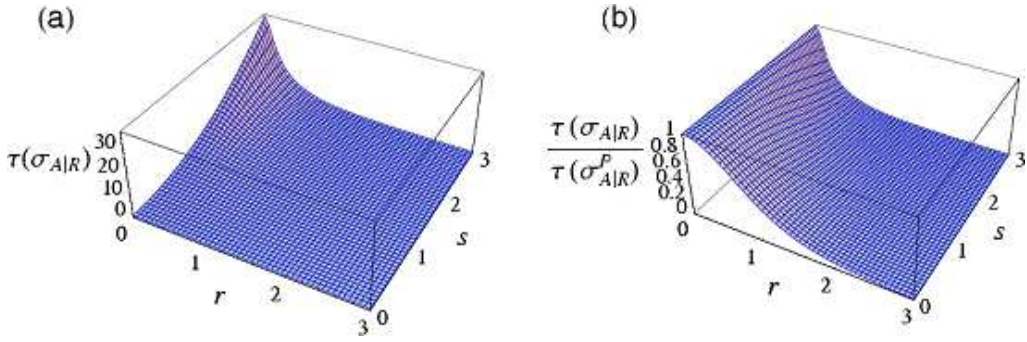}%
\caption{\label{fi3decay}(color online) Bipartite entanglement
described by Alice and the non-inertial observer Rob, who moves with
uniform acceleration parametrized by the effective squeezing $r$.
From an inertial perspective, the field is in a two-mode squeezed
state with squeezing degree $s$. Plot (a) depicts the contangle
$\tau(\sig_{A|R})$, given by Eqs.~(\ref{tau},\ref{m3_ar}), as a
function of $r$ and $s$. In plot (b) the same quantity is normalized
to the original contangle as seen by inertial observers,
$\tau(\sig^P_{A|R})=4s^2$. Notice in (a) how the bipartite contangle
is an increasing function of the entanglement, $s$, while it
decreases with increasing Rob's acceleration, $r$, vanishing in the
limit $r \rightarrow \infty$. This degradation is faster for higher
$s$, as clearly visible in (b).}}
\end{figure*}

Let us first comment on the quantum correlations created between the
two Rindler regions $I$ and $II$, given by \eq{m3_rr}. Note that the
entanglement in the mixed state $\sig_{R \bar R}$ is exactly equal,
in content, to that of a pure two-mode squeezed state with squeezing
$r$, unregardingly of the initial Alice-Rob entanglement quantified
by $s$. This provides a clearcut interpretation of the Unruh
mechanism, in which the acceleration alone is responsible of the
creation of entanglement between the accessible degrees of freedom
described by Rob, and the unaccessible ones described by the virtual
anti-Rob. By comparison with Ref.~\cite{ahn}, we remark that if the
logarithmic negativity is used as an entanglement measure, this
insightful picture is no longer true, as in that case the
entanglement described by Rob and anti-Rob depends on $s$ as well.
While this is not surprising given the aforementioned inequivalence
between negativities and Gaussian entanglement measures in
quantifying quantum correlation of nonsymmetric mixed Gaussian
states \cite{ordering}, it gives an indication that the negativity
is probably not the best quantifier to capture the transformation of
quantum information due to relativistic effects.

%Moreover, it follows as a strong conjecture from some results in
%bosonic channel capacity \cite{giovannetti, seramulti} that in
%two-mode GMEMMS obtained as reductions of a three-mode pure Gaussian
%state (as it is the case in our setting) the Gaussian entanglement
%of formation may coincide with the true entanglement of formation
% (which would be thus globally minimized over an
%ensemble of pure Gaussian states in this instance), providing the
%whole family of Gaussian entanglement measures with a stronger
%operational interpretation when applied to such states
%\cite{inprep}.

The proper quantification of Gaussian entanglement shows indeed that
the quantum correlations are regulated by two competing squeezing
degrees. One one hand, the resource parameter $s$ regulates the
entanglement
 $\tau(\sig^P_{A|R})=4s^2$ described by inertial
observers. On the other hand, the acceleration parameter $r$
regulates the uprising entanglement $\tau(\sig_{R|\bar R})=4s^2$
between the modes described by the uniformly accelerated Rob and by
his {\em alter ego} anti-Rob. The latter entanglement, obviously,
increases to the detriment of the entanglement
$\tau(\sig_{A|R})=g[m^2_{A|R}]$ described by Alice and Rob from the
non-inertial perspective. \eq{m3_ar} shows in fact that
$\tau(\sig_{A|R})$ is increasing with $s$ and decreasing with $r$,
as pictorially depicted in Fig. (\ref{fi3decay}). Interestingly,
 the rate at which this bipartite entanglement degrades with $r$,
$|\partial \tau(\sig_{A|R})/
\partial r|$, increases with $s$: for higher $s$ Alice and Rob
describe the field as more entangled (from the inertial perspective
which corresponds to $r=0$), but it drops faster when the
acceleration ($r$) comes into play. The same behavior is observed
for the negativity \cite{ahn}. For any inertial entanglement $s$, no
quantum correlations are left in the infinite acceleration limit ($r
\rightarrow \infty$), when the state $\sig_{A|R}$ becomes
asymptotically separable.

\begin{figure}[t!]
\centering{\includegraphics[width=8.5cm]{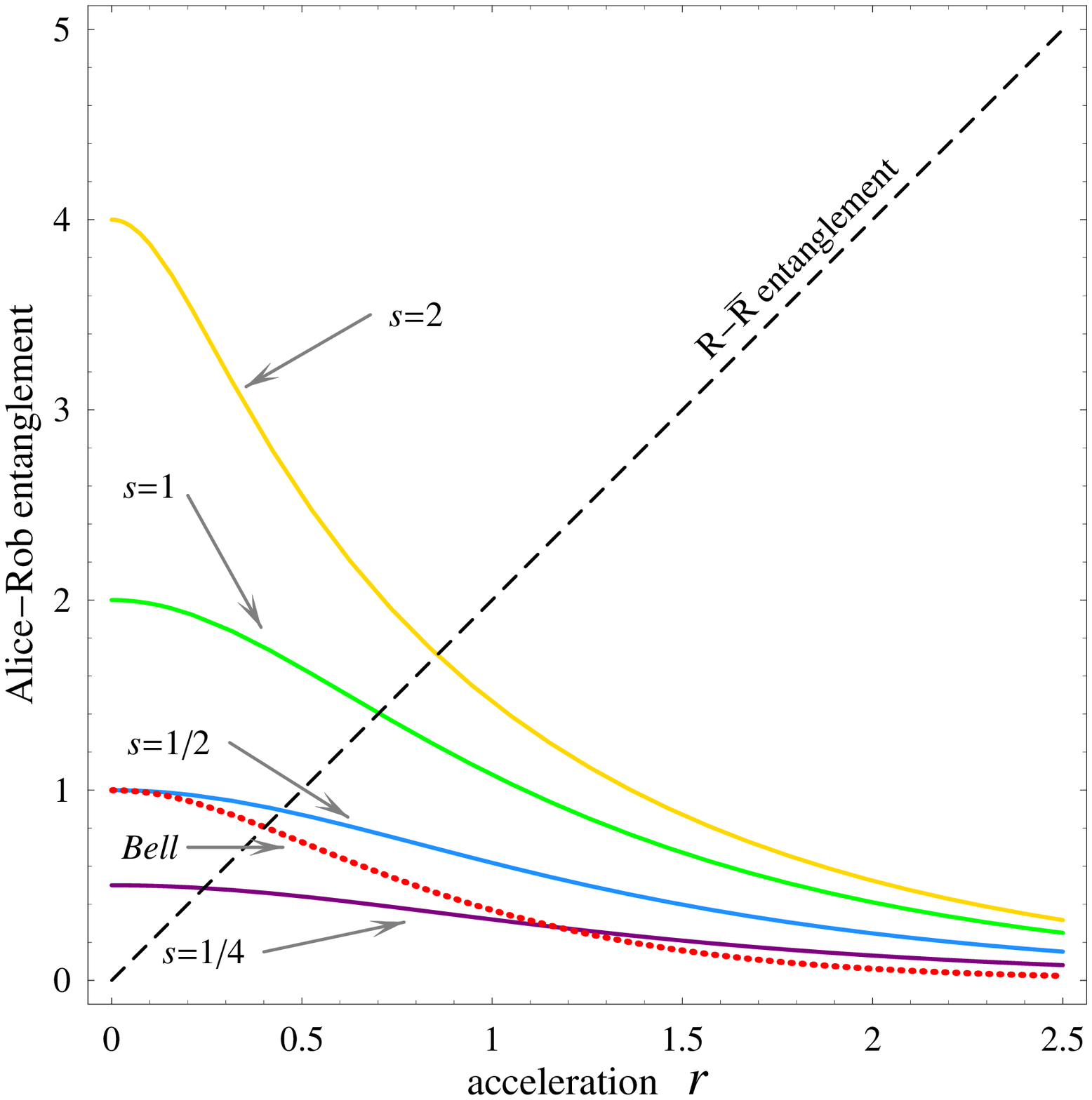}%
\caption{\label{comparob}(color online) Bipartite entanglement
between modes described by Alice and the non-inertial Rob moving
with uniform acceleration parametrized by $r$. The dotted red curve
depicts the dependence of the logarithmic negativity between modes
described by Alice and Rob in the instance of a state which
corresponds to a two-qubit Bell state from the inertial perspective
as computed in Ref.~\cite{alice}. The other solid curves correspond
to $\sqrt{\tau(\sig_{A|R})}$  (the square root of the contangle is
taken to provide a fair dimensional comparison) as computed in this
paper [see \eq{m3_ar}], in the instance of an entangled two-mode
squeezed state described from the perspective of two inertial
observers, with different squeezing parameters
$s=0.25,\,0.5,\,1,\,2$ (referring to the purple, blue, green and
gold curve, respectively). As a further comparison, the entanglement
described by Rob and anti-Rob, given by $\sqrt{\tau(\sig_{R|\bar
R})}=2r$ [see \eq{m3_rr}] independently of $s$, is plotted as well
(dashed black diagonal line).}}
\end{figure}

It is instructive to compare these results to the analysis of
entanglement when the field (for $r=0$) is in a two-qubit Bell state
$\sqrt{\frac12}\left(\ket{0}_{\alpha_M}\ket{0}_{\rho_M} +
\ket{1}_{\alpha_M}\ket{1}_{\rho_M}\right)$, where $\ket{1}$ stands
for the single-boson Fock state \cite{alice}. When one observer is
accelerated, the state belongs to a three-partite Hilbert space with
dimension $2\times\infty\times\infty$. The free entanglement in the
state is degraded with the acceleration and vanishes in the infinite
acceleration limit. Fig.~\ref{comparob} plots the entanglement
between modes described by Alice and the non-inertial Rob in such a
qubit-CV setting \cite{alice}, compared with the fully CV scenario
considered in this paper. When the field described from the inertial
perspective is in a two-mode squeezed Gaussian state with $s
> 1/2$, the entanglement is always stronger than the
entanglement in the Bell-state case. We also observe that, even for
$s< 1/2$, the degradation of entanglement with acceleration is slower for
the Gaussian state. The exploitation of all the infinitely-many
degrees of freedom available in the Hilbert space, therefore,
results in an improved robustness of the entanglement against the
thermalization induced by the Unruh effect.

In this context, we can pose the question of how much entanglement,
at most, can Alice and the non-inertial Rob hope to maintain, given
that Rob is moving with a finite, known acceleration $r$. Assuming
that from an inertial perspective the state is a perfect EPR state,
we find
\begin{equation}\label{limtau}
\lim_{s \rightarrow \infty} m_{A|R} = 1+2/\sinh^2(r)\,,
\end{equation}
meaning that the maximum entanglement left by the Unruh
thermalization, out of an initial unlimited entanglement, approaches
asymptotically
\begin{equation}\label{taumax}
\tau_r^{\max}(\sig_{A|R}) = {\rm
arcsinh}^2\left[\frac{2\cosh(r)}{\sinh^2(r)}\right]\,.
\end{equation}
Only for zero acceleration, $r=0$, this maximum entanglement
diverges. For any nonzero acceleration, the quantity
$\tau_r^{\max}(\sig_{A|R})$ is finite and rapidly degrades with $r$.
This provides an upper bound to the effective quantum correlations
and thus, the efficiency of any conceivable quantum information
protocol that Alice and the non-inertial Rob may implement. For
example, if Rob travels with a modest acceleration given by $r=0.5$,
no more than 8 ebits of entanglement are left between the modes
described by Alice and Rob, even if the state contained an
infinitely amount of entanglement from the point of view of inertial
observers. This apparent `loss' of quantum information will be
precisely understood in the next subsection, where we will show that
the initial bipartite entanglement does not disappear, but is
redistributed into tripartite correlations among Alice, Rob and
anti-Rob.

\subsection{Tripartite entanglement}

 A proper measure of genuine tripartite entanglement is available for
 any three-mode Gaussian state \cite{contangle,3modi}. The measure, known as
  the ``residual contangle'', emerges from the monogamy
inequality~(\ref{ckwine}) and is an entanglement monotone under
tripartite GLOCC for pure states. The residual contangle of a
three-mode ($i$, $j$, and $k$) Gaussian state $\sig$, is defined as
\cite{contangle}
\begin{equation}\label{taures}
\tau(\sig_{i|j|k})\equiv\min_{(i,j,k)} \left[
\tau(\sig_{i|(jk)})-\tau(\sig_{i|j})-\tau(\sig_{i|k})\right]\,,
\end{equation}
where $(i,j,k)$ denotes all the permutations of the three mode
indexes. For pure states, the minimum in \eq{taures} is always
attained by the decomposition realized with respect to the probe
mode $i$ with smallest local determinant
$\det{\sig_i}=m^2_{i|(jk)}$.

We can promptly apply such definition to compute the distributed
tripartite entanglement in the state $\sig_{AR \bar R}$. From
\eq{m3_12}, we find that $m_{\bar R|(AR)}<m_{ A|(R \bar R)}$ for
$r<r^\ast$, with
$$r^\ast = {\rm arccosh}\sqrt{\tanh ^2(s) + 1}\,,$$
 while $m_{R|(A \bar R)}$ is always bigger than the other two
quantities. Using
Eqs.~(\ref{tau},\ref{m3_12},\ref{m3_ar},\ref{m3_rr},\ref{taures})
together with $\tau(\sig_{A|\bar R})=0$, we find that the residual
contangle is given by
\begin{widetext}
\begin{eqnarray}\label{tau3} \hspace*{-2.5cm}
\tau(\sig_{A|R|\bar R}) &=& \left\{
                            \begin{array}{ll}
g[m^2_{\bar R|(AR)}]-g[m^2_{R|\bar R}], & \ r<r^\ast; \\
g[m^2_{A|(R \bar R)}]-g[m^2_{A|R}], & \hbox{otherwise.}
                            \end{array}
                          \right.   \\ \hspace*{-2.5cm}
&=&  \left\{
                            \begin{array}{ll}
                            - 4 r^2+{\rm arcsinh}^2 \sqrt{\left[\cosh ^2(r) + \cosh (2 s) \sinh ^2(r)\right]^2 \
- 1},  & \ r<r^\ast; \\
 4 s^2 - {\rm arcsinh}^2 \sqrt{\frac{\left[2 \sinh ^2(r) + (\cosh (2 r) +
                                        3) \cosh (2 s)\right]^2}{\left[2 \
\cosh (2 s) \sinh ^2(r) + \cosh (2 r) + 3\right]^2} - 1},
&\hbox{otherwise.} \nonumber
                            \end{array}
                          \right.
\end{eqnarray}
\end{widetext}

\begin{figure}[t!]
\centering{\includegraphics[width=8.5cm]{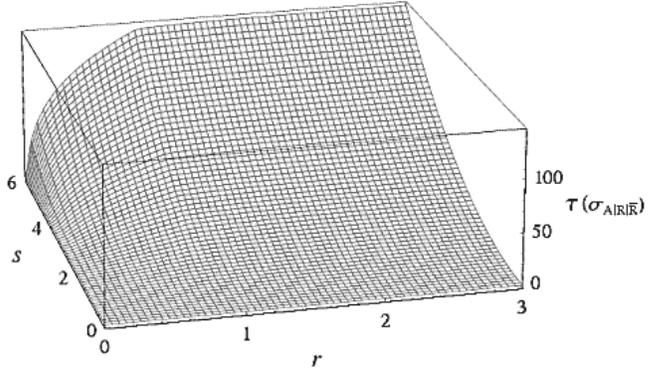}%
\caption{\label{tricont}(color online) Genuine tripartite
entanglement, as quantified by the residual contangle \eq{tau3},
among the inertial Alice, Rob in Rindler region $I$, and anti-Rob in
Rindler region $II$, plotted as a function of the initial squeezing
$s$ and of Rob's acceleration $r$. The tripartite entanglement
increases with $r$, and for ${r \rightarrow \infty}$ it approaches
the original entanglement content $4s^2$ between modes described by
Alice and Rob form the inertial perspective.}}
\end{figure}

The tripartite entanglement is plotted in Fig.~\ref{tricont} as a
function of $r$ and $s$. Very remarkably, for any initial squeezing
$s$ it increases with increasing acceleration $r$. In the limit of
infinite acceleration, the bipartite entanglement between modes
described by Alice and Rob vanishes so we have that
\begin{equation}\label{tau3asi}
\lim_{r \rightarrow \infty} \tau(\sig_{A|R|\bar R}) =
\tau(\sig_{A|(R \bar R)}) = \tau(\sig^P_{A|R})=4s^2\,.
\end{equation}
Precisely, {\em the genuine tripartite entanglement tends
asymptotically to the two-mode squeezed entanglement described by
inertial observers}.

We have now all the elements necessary to fully understand the Unruh
effect on CV entanglement of bosonic particles, when a single
observer is accelerated. Due to the fact that Rob is accelerated,
from his perspective:
\begin{itemize}
  \item there is bipartite entanglement between the two modes in the two distinct Rindler
regions, and this entanglement is only function of the acceleration;
  \item the bipartite entanglement described by inertial observers is
redistributed into a genuine tripartite entanglement among the modes
described by Alice, Rob and anti-Rob. Therefore, as a consequence of
the monogamy of entanglement, the entanglement between the two modes
described by Alice and Rob is degraded.
\end{itemize}

In fact, there is no bipartite entanglement between the modes
described by Alice and anti-Rob. This is very different to the
distribution of entanglement of Dirac fields from a non-inertial
perspective \cite{dirac}, where the fermionic statistics does not
allow the creation of maximal entanglement between the two Rindler
regions. Therefore, the entanglement between modes described by
Alice and Rob is never fully degraded. As a result of the monogamy
constraints on entanglement sharing \cite{ckw}, the mode described
by Alice becomes entangled with the mode described by anti-Rob and
the entanglement in the resulting three-qubit system is distributed
in couplewise correlations, and a genuine tripartite entanglement is
never created in that case \cite{dirac}.

In the next Section, we will show how in the bosonic case the
picture radically changes when both observers undergo uniform
acceleration, in which case the relativistic effects are even more
surprising.

\subsection{Mutual information}

\begin{figure*}[t]
\includegraphics[width=13cm]{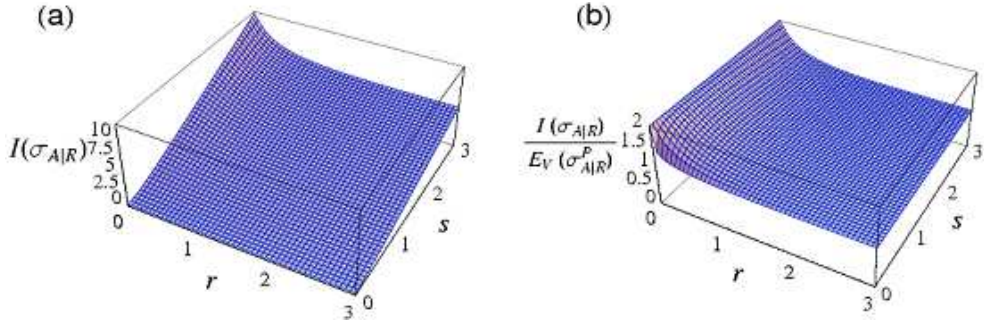}%
\caption{\label{mi3decay}(color online) Total correlations between
modes described by Alice and the non-inertial observer Rob, moving
with acceleration given by the effective squeezing parameter $r$.
From an inertial perspective, the field is in a two-mode squeezed
state with squeezing degree $s$. Plot (a) depicts the dependence of
the mutual information $I(\sig_{A|R})$, given by Eq.~(\ref{mi3}), as
a function of $r$ and $s$. In plot (b) the same quantity is
normalized to the entropy of entanglement as perceived by inertial
observers, $E_V(\sig^P_{A|R})$, \eq{ev3}. Notice in (a) how the
mutual information is an increasing function of the inertial
entanglement, $s$; at variance with the entanglement (see
Fig.~\ref{fi3decay}), it saturates to a nonzero value in the limit
of infinite acceleration. From plot (b), one clearly sees that this
asymptotic value is exactly equal to the  entropy of entanglement
described by inertial observers.}\end{figure*}

It is interesting to compute the total (classical and quantum)
correlations between modes described by Alice and the non-inertial
Rob, encoded in the reduced (mixed) two-mode state $\sig_{A|R}$ of
\eq{sig3}, using the mutual information $I(\sig_{A|R})$, \eq{migau}.
The symplectic spectrum of such state is constituted by
$\eta^-_{A|R}=1$ and $\eta^+_{A|R}=\sqrt{\det\sig_{\bar R}}$. Since
it belongs to the class of GMEMMS, it is in particular a mixed state
of partial minimum uncertainty, which saturates \ineq{bonfide}
\cite{extremal}. Therefore, the mutual information reads
\begin{equation}\label{mi3}
I(\sig_{A|R}) = f(\sqrt{\det\sig_A})+f(\sqrt{\det\sig_R}) -
f(\sqrt{\det\sig_{\bar R}})\,.
\end{equation}

Explicitly:\\
\begin{widetext}
\noindent $I(\sig_{A|R}) = \log [\cosh ^2(s) \sinh ^2(r)] \sinh
^2(r) \cosh ^2(s) + \log [\cosh ^2(s)] \cosh ^2(s) +  \log [\cosh
^2(r) \cosh ^2(s)] \cosh ^2(r) \cosh ^2(s) - \log [\sinh ^2(s)]
\sinh ^2(s) - \frac{1}{2} \log \{\frac{1}{2} [\cosh (2 s) \cosh
^2(r) + \sinh ^2(r) - 1]\} [\cosh (2 s) \cosh ^2(r) + \sinh ^2(r) -
        1] - \frac{1}{2} \log \{\frac{1}{2} [\cosh ^2(r) + \cosh (2 s) \sinh ^2(r) +
              1]\} [\cosh ^2(r) + \cosh (2 s) \sinh ^2(r) +
        1]$.
\end{widetext}

The mutual information of \eq{mi3} is plotted in
Fig.~\ref{mi3decay}(a) as a function of the squeezing degrees $s$
(corresponding to the entanglement described from the inertial
perspective) and $r$ (reflecting Rob's acceleration). It is
interesting to compare the mutual information with the original
two-mode squeezed entanglement described between the inertial
observers. In this case, it is more appropriate to quantify the
entanglement in terms of the entropy of entanglement,
$E_V(\sig_{A|R}^P$), defined as the Von Neumann entropy of each
reduced single-mode CM, $E_V(\sig_{A|R}^P) \equiv
S_V(\sig_A^P)\equiv S_V(\sig_B^P)$. Namely,
\begin{equation}\label{ev3}
E_V(\sig_{A|R}^P)= f(\cosh 2s)\,,
\end{equation}
with $f(x)$ given by \eq{entfunc}. From the perspective of inertial
observers ($r=0$), the state is pure, $\sig_{A|R}\equiv\sig_{A|R}^P$
and the mutual information is equal to twice the entropy of
entanglement of \eq{ev3}, meaning that the two modes described by
inertial observers are correlated both quantumly and classically to
the same degree. When Rob is under acceleration ($r \neq 0$), the
entanglement with the modes described by Alice is degraded by the
Unruh effect (see Fig.~\ref{fi3decay}), but the classical
correlations are left untouched. In the limit $r \rightarrow
\infty$, all entanglement is destroyed and the remaining mutual
information $I(\sig_{A|R})$, quantifying classical correlations
only, saturates to $E_V(\sig_{A|R}^P)$ from \eq{ev3}. For any $s>0$
the mutual information of \eq{mi3}, once normalized by such entropy
of entanglement [see Fig.~\ref{mi3decay}(b)], ranges between $2$
($1$ normalized unit of entanglement plus $1$ normalized unit of
classical correlations) at $r=0$, and $1$ (all classical
correlations and zero entanglement) at $r \rightarrow \infty$. The
same behavior is found for classical correlations in the case of
 entangled states of the field which are bosonic two-qubit Bell states in an
inertial perspective \cite{alice}.

\section{Distributed Gaussian entanglement due to both accelerated
observers} \label{secTwo}

A natural question arises whether the mechanism of degradation or,
to be precise, distribution of entanglement due to the Unruh effect
is qualitatively modified according to the number of accelerated
observers. One might guess that when both observers travel with
uniform acceleration, basically the same features as unveiled above
for the case of a single non-inertial observer will manifest, with a
merely quantitative rescaling of the relevant figures of merit (such
as bipartite entanglement degradation rate). Indeed, we will now
show that this is {\em not} the case.

\begin{figure}[t!]
\centering{\includegraphics[width=7cm]{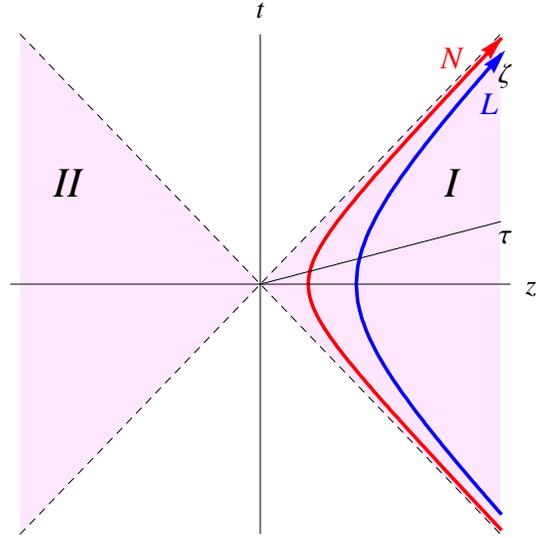}%
\caption{\label{mondodue}(color online) Sketch of the world lines
for the two non-inertial observers Leo and Nadia. The set $(z,t)$
denotes Minkowski coordinates, while the set $(\zeta, \tau)$ denotes
Rindler coordinates. The causally disconnected Rindler regions $I$
and $II$ are evidenced.}}
\end{figure}

We consider here two non-inertial observers, with different names
for ease of clarity and to avoid confusion with the previous
picture. Leo and Nadia both travel with uniform accelerations
$\aleph_L$ and $\aleph_N$, respectively and describe the state of a
scalar field in Rindler coordinates from their non-inertial
perspective (see Fig.~\ref{mondodue}). Similarly to the previous
instance, we consider that from the perspective of inertial
observers only two field modes, of frequencies $\lambda$ and $\nu$,
are entangled in a pure two-mode squeezed state $\sig^{P}_{LN}(s)$
of the form \eq{tms}, with squeezing parameter $s$ as before. Due to
the acceleration of both observers, the entanglement is
redistributed among modes described by four observers: Leo, Nadia
(living in Rindler region $I$), anti-Leo and anti-Nadia (living in
Rindler region $II$). These four (some real and some virtual)
parties will describe modes $\lambda_{I}$, $\nu_I$, $\lambda_{II}$,
$\nu_{II}$, respectively. By the same argument of Sec.~\ref{secOne},
the four observers will describe a pure four-mode Gaussian state
with CM given by \cite{noteid}
\begin{equation}\label{in4}
\begin{split}
\sig_{\bar L L N \bar N}(s,l,n) &= S_{\lambda_I,\lambda_{II}}(l)
S_{\nu_I,\nu_{II}}(n) S_{\lambda_I,\nu_I}(s)\cdot \id_{8} \\ &\
\cdot\ S_{\lambda_I,\nu_I}^T(s) S_{\nu_I,\nu_{II}}^T(n)
S_{\lambda_I,\lambda_{II}}^T(l)\,, \end{split}
\end{equation}
where the symplectic transformations $S$ are given by \eq{tmsS},
 $\id_{8}$ is the CM of the vacuum
$\ket{0}_{\lambda_{II}}\!\otimes\!\ket{0}_{\lambda_{I}}\!\otimes\!\ket{0}_{\nu_I}\!\otimes\!\ket{0}_{\nu_{II}}$,
and $l$ and $n$ are the squeezing parameters associated with the
respective  accelerations $\aleph_L$ and $\aleph_N$ of Leo and Nadia
[see \eq{accparam}]. Explicitly,
\begin{equation}\label{sig4}
\sig_{\bar L L N \bar N}={\left(%
 \begin{array}{cccc}
  \sig_{\bar L} & \eps_{\bar L L} &  \eps_{\bar L N}  & \eps_{\bar L \bar N}\\
  \eps^T_{\bar L L} & \sig_L & \eps_{L N} & \eps_{L \bar N}\\
  \eps^T_{\bar L N}  & \eps^T_{L N} & \sig_{N} & \eps_{N \bar N} \\
  \eps^T_{\bar L \bar N} & \eps^T_{L \bar n} & \eps^T_{N \bar N} &
  \sig_{\bar N}
\end{array}%
\right)}\,,
\end{equation}
where:
\begin{eqnarray*}
\sig_{\bar X} &=& [\cosh^2(x) + \cosh (2 s) \sinh^2(x)] \id_2\,, \\
 \sig_{X} &=& [\cosh^2(x)\cosh(2s)+\sinh^2(x)] \id_2\,, \\
\eps_{\bar X X}=\eps_{X \bar X} &=& [\cosh^2(s) \sinh (2 x)]
Z_2\,,\\
\eps_{\bar X Y}=\eps_{Y \bar X} &=& [\cosh (y) \sinh (2 s) \sinh
(x)] \id_2 \,, \\
\eps_{\bar X \bar Y}&=&[\sinh (2 s) \sinh (x) \sinh (y)] Z_2\,,\\
\eps_{X Y} &=& [\cosh (x) \cosh (y) \sinh (2 s)] Z_2\,,
\end{eqnarray*}
\\ with $Z_2={{1\ \ \ 0}\choose {0 \ -1}}$;  $X,Y=\{L,N\}$ with $X
\ne Y$,  and
 accordingly for the lower-case symbols $x,y=\{l,n\}$.

The very high acceleration regime ($l,n \gg 0$) can now be
interpreted as Leo and Nadia both escaping the fall into the black
hole by accelerating away from it with acceleration $\aleph_L$ and
$\aleph_N$, respectively. Their entanglement will be degraded since
part of the information is lost through the horizon into the black
hole. Their acceleration makes part of the information unavailable
to them. We will show that this loss involves both quantum and
classical information.

\subsection{Bipartite entanglement}
We first recall that the original contangle
$\tau(\sig^P_{L|N})=4s^2$ described by two inertial observers is
preserved under the form of bipartite four-mode entanglement
$\tau(\sig_{(\bar L L)|(N \bar N)})$ between the two real and the
two virtual observers, as the two Rindler change of coordinates
amount to local unitary operations with respect to the $(\bar L
L)|(N \bar N)$ bipartition. The computation of the bipartite
contangle in the various $1 \times 1$ partitions of the state
$\sig_{\bar L L N \bar N}$ is still possible in closed form thanks
to the results of Ref.~\cite{ordering}. From
Eqs.~(\ref{tau},\ref{sig4}), we find
\begin{widetext}
\begin{eqnarray}
 m_{L|\bar N} &=& m_{N|\bar L}\,=\,m_{\bar L|\bar
N}\,=\,1\,,   \label{m411sep} \\
 m_{L|\bar L} &=& \cosh(2l)\,,\quad m_{N|\bar N}\,=\,\cosh(2n)\,, \label{m4xbarx}\\
  m_{L|N} &=& \left\{
                \begin{array}{ll}
                  1\,, & \ \ \tanh (s)\leq \sinh (l) \sinh (n)\,; \\
                  \frac{2 \cosh (2 l) \cosh (2 n) \cosh ^2(s) + 3 \cosh (2 s) -
      4 \sinh (l) \sinh (n) \sinh (2 s) -
      1}{2 \left[(\cosh (2 l) + \cosh (2 n)) \cosh ^2(s) - 2 \sinh ^2(s) +
          2 \sinh (l) \sinh (n) \sinh (2 s)\right]}, &
          \ \ \hbox{otherwise}\,.
                \end{array}
              \right. \label{m4ln}
\end{eqnarray}
\end{widetext}

Let us first comment on the similarities with the setting of an
inertial Alice and a non-inertial Rob. In the case of two
accelerated observers, \eq{m411sep} entails (we remind that $m=1$
means separability) that the mode described by Leo (Nadia) never
gets entangled to the mode described by anti-Nadia (anti-Leo).
Naturally, there is no bipartite entanglement generated between the
modes described by the two virtual observers $\bar L$ and $\bar N$.
Another similarity found in \eq{m4xbarx}, is that the reduced
two-mode state $\sig_{X \bar X}$ assigned to each observer
$X=\{L,N\}$ and her/his respective virtual counterpart $\bar X$, is
exactly of the same form as $\sig_{R \bar R}$. Therefore, due to the
fact that Leo and Nadia are accelerated, from their perspective we
find again that a bipartite contangle is present between the mode
described by each observer in region $I$, and the corresponding
causally disconnected mode described by the respective {\em alter
ego} virtual observer in region $II$; this entanglement is a
function of the corresponding acceleration $x=\{l,n\}$ only. The two
entanglements corresponding to each observer--anti-observer pair are
mutually independent, and for each the $X|\bar X$ entanglement
content is again the same as that of a pure, two-mode squeezed state
created with squeezing parameter $x$.

The only entanglement which is physically accessible to the
non-inertial observers is encoded in the two modes $\lambda_I$ and
$\nu_I$ corresponding to Rindler regions $I$ of Leo and Nadia. These
two modes are left in the state $\sig_{LN}$, which is not a GMEMMS
(like the state $\sig_{AR}$ in Sec.~\ref{secOne}) but a nonsymmetric
thermal squeezed state \cite{extremal}, for which the Gaussian
entanglement measures are available as well \cite{ordering}. The
contangle of such state is in fact given by \eq{m4ln}. Here we find
a first significant qualitative difference with the case of a single
accelerated observer: a state entangled from an inertial perspective
can become disentangled for two non-inertial observers, both
traveling with {\em finite} acceleration. \eq{m4ln} shows that there
is a trade-off between the amount of entanglement ($s$) described
form an inertial perspective, and the acceleration parameters of
both parties ($l$ and $n$). If the observers are highly accelerated
(namely, if $\sinh (l) \sinh (n)$ exceeds $\tanh (s)$), the
entanglement in the state $\sig_{LN}$ vanishes, or better said,
becomes physically unaccessible to the non-inertial observers. Even
in the ideal case, where the state contains infinite entanglement
(corresponding to $s \rightarrow \infty$) from the perspective of
inertial observers, the entanglement {\em completely} vanishes from
the perspective of one inertial and one non-inertial observer if
$\sinh (l) \sinh (n) \ge 1$. We find here another important
difference to the Dirac case where entanglement never vanishes for
two non-inertial observers \cite{dirac}. Conversely, for any
nonzero, arbitrarily small acceleration parameters $l$ and $n$,
there is a threshold on the entanglement $s$ such that, if the
entanglement is smaller than the threshold, it vanishes when
described from the perspective of one inertial and one non-inertial
observer. With only one non-inertial observer, instead
(Sec.~\ref{secOne}), any infinitesimal entanglement will survive for
arbitrarily large acceleration, vanishing only in the infinite
acceleration limit.

To provide a better comparison between the two settings, let us
address the following question. Can the entanglement degradation
observed by Leo and Nadia (with acceleration parameters $l$ and $n$
respectively) be observed by an inertial Alice and a non-inertial
Rob traveling with some effective acceleration $r^{eff}$? We will
look for a value of $r^{eff}$ such that the reduced state
$\sig_{AR}$ of the three-mode state in \eq{sig3} is as entangled as
the reduced state $\sig_{LN}$ of the four-mode state in \eq{sig4}.
The problem can be straightforwardly solved by equating the
corresponding contangles \eq{m3_ar} and \eq{m4ln}, to obtain
\begin{equation}\label{reff}
r^{eff}=\left\{
  \begin{array}{ll}
    {\rm arccosh}\left[\frac{\cosh (l) \cosh (n) \sinh (s)}{\sinh (s) - \cosh (s) \sinh \
(l) \sinh (n)}\right], & \\ & \hspace*{-3cm} \tanh (s)> \sinh (l)
\sinh (n)\,; \\ &  \\
    \infty, & \hspace*{-.5cm}\hbox{otherwise.}
  \end{array}
\right.
\end{equation}
Clearly, for very high acceleration parameters $l$ and $n$ (or,
equivalently, very small inertial entanglement $s$) the information
loss due to the non-inertiality of both observers is only matched by
an infinite effective acceleration in the case of a single
non-inertial observer. In the regime in which entanglement does not
completely vanish, the effective acceleration of Rob in the
equivalent single--non-inertial--observer setting is a function of
the inertial entanglement $s$, as well as of the accelerations of
Leo and Nadia.

\subsubsection{Entanglement between different frequency modes}

\begin{figure*}[t]
\centering{\includegraphics[width=13cm]{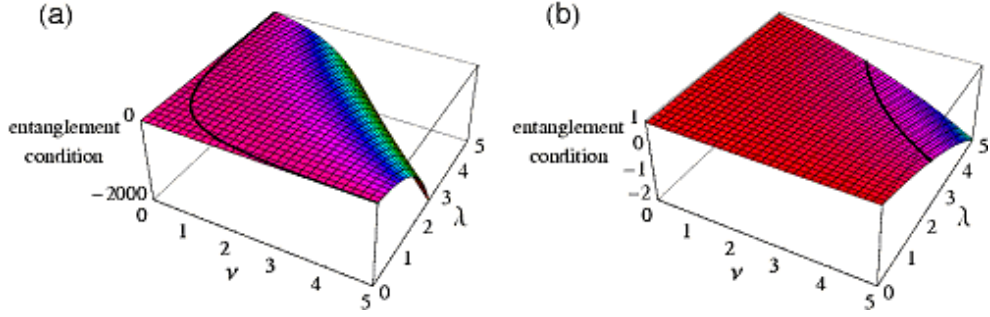}%
\caption{\label{condition}(color online) Entanglement condition,
\ineq{freqcondom}, for different frequency modes assuming that Leo
and Nadia have the same acceleration a) ${\aleph}=2\pi$ and b)
${\aleph}=10\pi$. Entanglement is only present in the frequency
range where the plotted surfaces assume negative values, and
vanishes for frequencies where the plots become positive; the
threshold [saturation of \ineq{freqcondom}] is highlighted with a
black line. Only modes whose frequencies are sufficiently high
exhibit bipartite entanglement. For higher accelerations of the
observers, the range of entangled frequency modes gets narrower, and
in the infinite acceleration limit the bipartite entanglement
between all frequency modes vanish.}}
\end{figure*}

The condition on the acceleration parameters $l$ and $n$ for which
the entanglement of the maximally entangled state ($s \rightarrow
\infty$) vanishes, from \eq{m4ln}, corresponds to the following
condition
$e^{\pi\Omega_{L}}+e^{\pi\Omega_{N}}-e^{\pi(\Omega_{L}+\Omega_{N})}\geq
0 $ where $\Omega_{L}=2\lambda/(\aleph_{L})$ and
$\Omega_{N}=2\nu/(\aleph_{N})$. Here we recall that $\aleph_{L,N}$
are the proper accelerations of the two non-inertial observers and
$\lambda,\nu$ the frequencies of the respective modes, see
\eq{accparam}. We assume now that Leo and Nadia have the same
acceleration, $$\aleph_{L}=\aleph_{N}\equiv {\aleph}$$ and ask the
question of, given their acceleration, which frequency modes would
they describe as entangled. This provides a deeper understanding of
the effect of the Unruh thermalization on the distribution of CV
correlations.

Our results immediately show that in this context the entanglement
{\em vanishes} between field modes such that
\begin{equation}
\label{freqcondom} e^{\frac{2\pi}{{\aleph}}\lambda}+e^{\frac{2\pi
}{{\aleph}}\mu}-e^{\frac{2\pi}{{\aleph}}(\lambda+\nu)}\geq 0\,.
\end{equation}
This means that if the field is, from the inertial perspective, in a
two-mode squeezed state with frequencies satisfying \eq{freqcondom},
the accelerated Leo and Nadia would describe the field as in a
two-mode disentangled (separable) state.  We have thus a practical
condition to determine which modes would be entangled from Leo and
Nadia's non-inertial perspective, depending on their frequency.

In Fig.~\ref{condition} we plot the condition on entanglement for
different frequency modes. The modes become disentangled when the
graph takes positive values. We see that only modes with the highest
frequencies exhibit bipartite entanglement for a given acceleration
${\aleph}$ of the observers. The larger the acceleration the less
modes remain entangled, as expected. In the limit of infinite
acceleration $\lambda/(\aleph_{L}),\nu/(\aleph_{N}) \rightarrow
\infty$, the set of entangled modes becomes empty. In the high
acceleration regime, where Alice and Rob escape the fall into a
black hole only very high frequency modes remain entangled.

\begin{figure}[t]
\centering{\includegraphics[width=8cm]{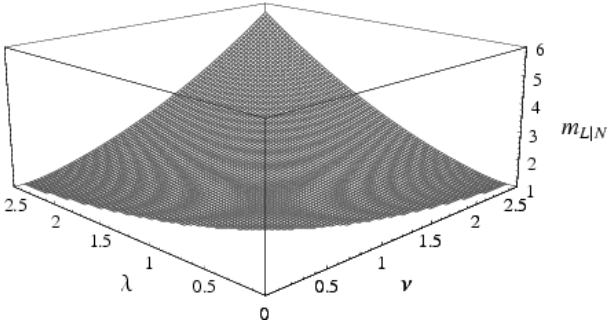}%
\caption{\label{entfrq}(color online) Entanglement between different
frequency modes assuming that Leo and Nadia have the same
acceleration ${\aleph}=2\pi$.  }} \end{figure}

Considering once more equally accelerated observers,
$\aleph_{L}=\aleph_{N}\equiv {\aleph}$ with finite ${\aleph}$, it is
straightforward to compute the contangle of the modes that do remain
entangled, in the case of a state which is maximally (infinitely)
entangled from the inertial perspective. From \eq{m4ln}, we have
\begin{equation}\label{entlims}
 m_{L|N}(s\rightarrow\infty)=\frac{\cosh (2 l) \cosh (2 n)-4
   \sinh (l) \sinh (n)+3}{2
   [\sinh (l)+\sinh (n)]^2}\,.
\end{equation}
In Fig.~\ref{entfrq} we plot the entanglement between the modes,
\eq{entlims}, as a function of their frequency $\lambda$ and $\nu$
[using \eq{accparam}] when Leo and Nadia have acceleration ${\aleph}
= 2\pi$. We see that, consistently with the previous analysis, at
fixed a acceleration, the entanglement is larger for higher
frequencies.  In the infinite acceleration limit, as already
remarked, entanglement vanishes for all frequency modes.

\begin{figure*}[t!]
\centering{\includegraphics[width=13cm]{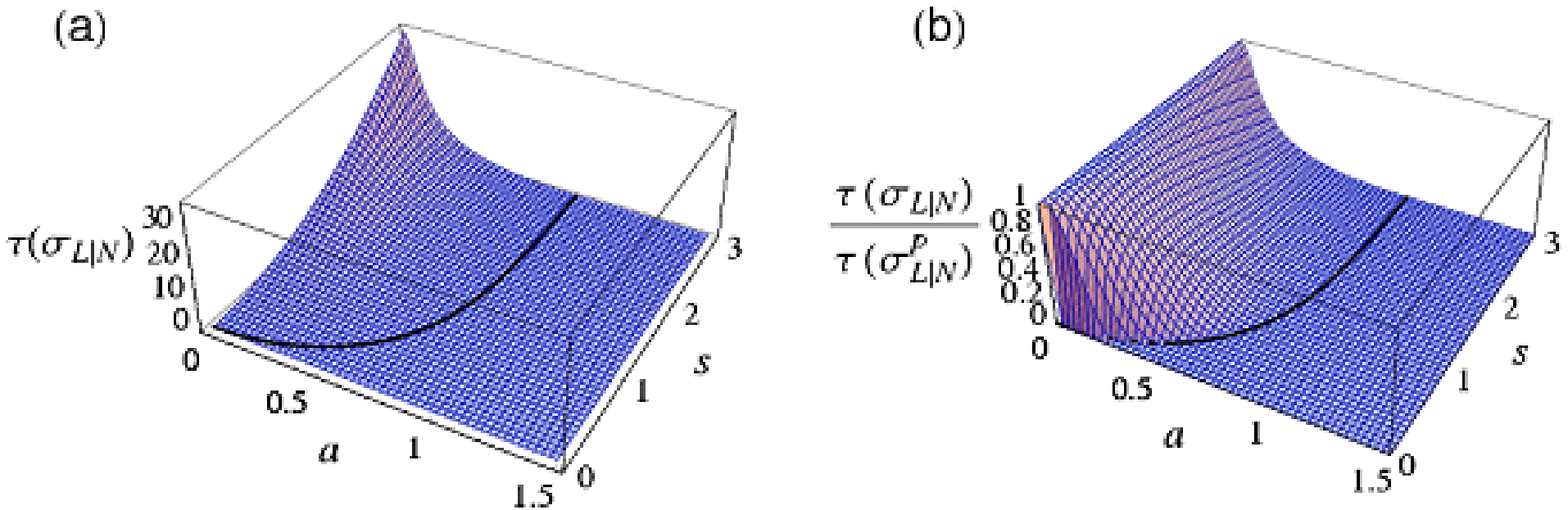}%
\caption{\label{fi4decay}(color online) Bipartite entanglement
between the modes described by the two non-inertial observers Leo
and Nadia, both traveling with uniform acceleration given by the
effective squeezing parameter $a$. From an inertial perspective, the
field is in a two-mode squeezed state with squeezing degree $s$.
Plot (a) depicts the contangle $\tau(\sig_{L|N})$, given by
Eqs.~(\ref{tau},\ref{m4lnaa}), as a function of $a$ and $s$. In plot
(b) the same quantity is normalized to the inertial contangle,
$\tau(\sig^P_{L|N})=4s^2$. Notice in (a) how the bipartite contangle
is an increasing function of the inertial entanglement, $s$, while
it decreases with increasing acceleration, $a$. This degradation is
faster for higher $s$, as clearly visible in (b). At variance with
the case of only one accelerated observer [Fig.~\ref{fi3decay}], in
this case the bipartite entanglement can be completely destroyed at
finite acceleration. The black line depicts the threshold
acceleration $a^\ast(s)$, \eq{ast}, such that for $a \ge a^\ast(s)$
the bipartite entanglement described by the two non-inertial
observers is exactly zero.}}
\end{figure*}

\subsubsection{Equal acceleration parameters}

We proceed now to analyze the bipartite entanglement in the field
from the non-inertial perspective, at fixed mode frequencies. For
simplicity, we restrict our attention to the case where Leo and
Nadia's trajectories have the same acceleration parameter
\begin{equation}\label{equa}
l=n\equiv a\,.
\end{equation}
This means that $\lambda/\aleph_{L}=\nu/\aleph_{N}$. While the
following results do not rely on this assumption, it is particularly
useful in order to provide a pictorial representation of
entanglement in the four-mode state $\sig_{\bar L L N \bar N}$,
which is now parameterized only by the two competing squeezing
degrees, the inertial quantum correlations ($s$) and the
acceleration parameter of both observers ($a$). In this case, the
acceleration parameter $a^\ast$ for which the entanglement between
the modes described by Leo and Nadia vanishes, is
\begin{equation}\label{ast}
a^\ast(s) = {\rm arcsinh} \left[\sqrt{\tanh(s)}\right]\,,
\end{equation}
where we used \eq{m4ln}. The contangle in the state $\sig_{LN}$ is
therefore given by
\begin{equation}\label{m4lnaa}
m_{L|N} = \left\{
                \begin{array}{l}
                  1\,, \hspace*{0.6cm} a \ge a^\ast(s)\,; \\ \\
                  \frac{2 \cosh ^2(2 a) \cosh ^2(s) + 3 \cosh (2 s) -
      4 \sinh ^2(a) \sinh (2 s) -
      1}{4 \left[\cosh ^2(a) + e^{2 s} \sinh ^2(a)\right]}
      ,  \\ \hspace*{1cm}
          \hbox{otherwise}\:\!,
                \end{array}
              \right.
\end{equation}
which we plot in Fig.~\ref{fi4decay}. The entanglement increases
with $s$ and decreasing with $a$ with a stronger rate of degradation
 for increasing $s$. The main difference with Fig.~\ref{fi3decay} is that entanglement here completely vanishes at
finite acceleration. Even if the state contains infinite
entanglement as described by inertial observers, entanglement
vanishes at $a \ge {\rm arcsinh} (1) \approx 0.8814$.

\subsection{Residual multipartite entanglement} \label{secTwoMulti}
It is straightforward to show that the four-mode state $\sig_{\bar L
L N \bar N}$ of \eq{sig4} is fully inseparable, which means that it
contains multipartite entanglement distributed among all the four
parties involved. This follows from the observation that the
determinant of each reduced one- and two-mode CM obtainable from
$\sig_{\bar L L N \bar N}$ is strictly bigger than $1$ for any
nonzero squeezings. This in addition to the global purity of the
state means that there is entanglement across all global
bipartitions of the four modes. We now aim to provide a quantitative
characterization of such multipartite entanglement.

For ease of simplicity, we focus once more on the case of two
observers with equal acceleration parameter $a$. The state under
consideration is obtained from \eq{sig4} via the prescription
\eq{equa}. The entanglement properties of this four-mode pure
Gaussian state have been investigated in detail by some of us
\cite{unlim}. We showed, in particular, that the entanglement
sharing structure in such state is infinitely {\em promiscuous}. The
state admits  the coexistence of an unlimited, genuine four-partite
entanglement, together with an accordingly unlimited bipartite
entanglement in the reduced two-mode states of two pair of parties,
here referred to as \{Leo, anti-Leo\}, and \{Nadia, anti-Nadia\}.
Both four-partite and bipartite correlations increase with $a$. We
will now review the study of multipartite entanglement of these
four-partite Gaussian states shown in Ref.~\cite{unlim}, with the
particular aim of showing the effects of relativistic acceleration
in the distribution of quantum information.

\subsubsection{Monogamy inequality} We begin by verifying that the state $\sig_{\bar L L N \bar N}$
satisfies the fundamental monogamy inequality (\ref{ckwine}) for the
entanglement distributed among the four parties (each one describing
a single mode). To this aim, we compute the pure-state contangle
between one probe mode and the remaining three modes. From \eq{tau}
we find
\begin{eqnarray}\label{m41x3}
  m_{\bar L\vert (L N \bar N)} =m_{\bar N\vert (NL \bar L)} &=&\cosh ^2a+\cosh (2s)\sinh ^2a \,,\nonumber\\
  m_{L\vert (\bar L N \bar N)} =m_{N\vert (\bar N L \bar L)} &=& \sinh ^2a+\cosh (2s)\cosh
^2a\,. \nonumber \\
\end{eqnarray}
Thanks to the explicit expressions
Eqs.~(\ref{m411sep}--\ref{m4ln},\ref{m41x3}) for the bipartite
entanglements, proving monogamy reduces to showing that $\min
\{g[m_{\bar L\vert (L N bar N)}^2 ]-g[m_{\bar L\vert L}^2
],\,g[m_{L\vert (\bar L N \bar N)}^2 ]-g[m_{\bar L\vert L}^2
]-g[m_{L\vert N}^2 ]\}$ is nonnegative. One can verify that the
first quantity always achieves the minimum, therefore we define
\begin{eqnarray}\label{tau4}
&&\tau^{res}(\sig_{\bar L L N \bar N})\equiv \tau (\sig_{\bar L | (L
N \bar N}))-\tau (\sig_{\bar L|L})\nonumber \\
&&={\rm arcsinh}^2\left\{ {\sqrt {[\cosh ^2a+\cosh (2s)\sinh
^2a]^2-1} } \right\}-4a^2 \nonumber \\ &&>0\,.
\end{eqnarray}
The residual contangle $\tau^{res}$ is positive as $\cosh (2s)>1$
for $s>0$, and it quantifies precisely the multipartite correlations
that cannot be stored in bipartite form. Those quantum correlations,
however, can be either tripartite involving three of the four modes,
and/or genuinely four-partite among all of them. We can now
quantitatively estimate to what extent such correlations are encoded
in some tripartite form: as an anticipation, we will find them
negligible in the limit of high acceleration.

\subsubsection{Tripartite entanglement.}

Let us first observe that in the tripartitions $\bar L | L | \bar N$
and $\bar L | N | \bar N$ the tripartite entanglement is exactly
zero. This is because the mode described by anti-Nadia is not
entangled with the modes described by the pair \{Leo, anti-Leo\},
and the mode described by anti-Leo is not entangled with the modes
described by the pair \{Nadia, anti-Nadia\}. The corresponding
three-mode states are then said to be biseparable \cite{barbarella}.
The only tripartite entanglement present, if any, is equal in
content (due to the symmetry of the state) for the tripartitions
$\bar L | L | N$ and $L |N|\bar N$. It is properly quantified by the
residual tripartite contangle $\tau(\sig_{\bar L | L | N})$ emerging
from the corresponding mixed-state three-mode monogamy inequality,
via \eq{taures}. In Ref.~\cite{unlim} an upper bound on
$\tau(\sig_{\bar L | L | N})$ has been obtained. Its derivation is
recalled in Appendix \ref{secApp} for the sake of completeness. From
\eq{tribnd} we have
\begin{eqnarray}\label{tau43}
&&\hspace*{-0.5cm}\tau(\sig_{\bar L | L | N}) \le
\tau^{bound}(\sig_{\bar L | L |
N}) \\
\nonumber &&\hspace*{-0.5cm}\equiv \min \Bigg\{g\left[\left(
\cosh^2a+\frac{1+{\rm sech}^2a\tanh^2s} {1-{\rm sech}^2a\tanh^2s}
\sinh^2a
\right)^2\right]\!-4a^2,\,\\
\nonumber &&\hspace*{.8cm} g\left[\left(\frac{1+{\rm
sech}^2a\tanh^2s} {1-{\rm
sech}^2a\tanh^2s}\right)^2\right]-g\left[m_{L\vert N}^2
\right]\Bigg\}\,,
\end{eqnarray}
with $m_{L\vert N}$ obtainable by substituting \eq{equa} in
\eq{m4ln}.

The upper bound $\tau^{bound}(\sig_{\bar L | L | N})$ is of course
always nonnegative (as a consequence of monogamy), it decreases with
increasing acceleration $a$, and vanishes in the limit $a\to \infty
$. Therefore, in the regime of increasingly high $a$, eventually
approaching infinity, any form of tripartite entanglement among any
three modes in the state $\sig_{\bar L L N \bar N}$ is negligible
(exactly zero in the limit of infinite acceleration).

\subsubsection{Genuine four-partite entanglement.}

\begin{figure}[t!]\centering{\includegraphics[width=8.5cm]{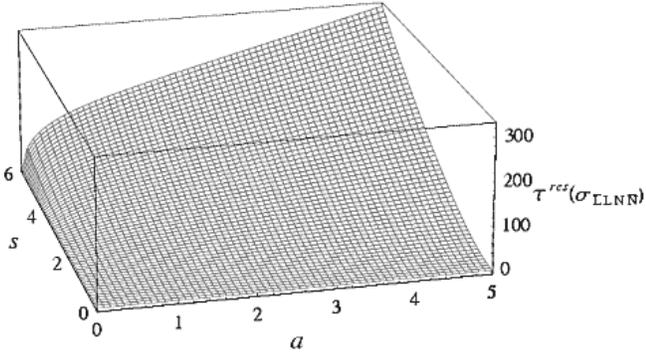}%
\caption{\label{figfour}(color online) Residual contangle,
\eq{tau4}, not stored in bipartite form, distributed among the modes
described by the non-inertial observers Rob and Nadia in Rindler
region $I$ and their virtual counterparts anti-Leo and anti-Nadia in
Rindler region $II$,  as quantified by the residual contangle
\eq{tau3}, among the inertial Alice,  plotted as a function of the
initial squeezing $s$ and of the acceleration $a$ of both observers.
In the regime of high acceleration ($a \rightarrow \infty$), the
displayed residual entanglement is completely distributed in the
form of genuine four-partite quantum correlations. This four-partite
entanglement is monotonically increasing with increasing
acceleration $a$, and diverges as $a$ approaches infinity.}}
\end{figure}

The above analysis of the tripartite contribution to multipartite
entanglement shows that, in the regime of high acceleration $a$, the
residual entanglement $\tau^{res}$ determined by \eq{tau4} is stored
entirely in the form of four-partite quantum correlations.
Therefore, the residual entanglement in this case is a good measure
of \textit{genuine} four-partite entanglement among the four Rindler
spacetime modes. It is now straightforward to see that
$\tau^{res}(\sig_{\bar L L N \bar N})$ is itself an {\em increasing}
function of $a$ for any value of $s$ (see Fig.~\ref{figfour}), and
it \textit{diverges} in the limit $a\to \infty $.

The four-mode state \eq{tau4} obtained with an arbitrarily large
acceleration $a$, consequently, exhibits a coexistence of unlimited
genuine four-partite entanglement, and pairwise bipartite
entanglement in the reduced two-mode states $\sig_{L|\bar L}$ and
$\sig_{N | \bar N}$. This peculiar distribution of CV entanglement
 in the considered Gaussian state has been
defined as {\em infinitely promiscuous} in Ref.~\cite{unlim}. The
properties of such entangled states are discussed in
Ref.~\cite{unlim} in a practical optical setting. It is interesting
to note that in the relativistic analysis we present here,  the
genuine four-partite entanglement increases unboundedly with the
observers' acceleration. This is in fact in strong contrast with the
case of an inertial observer and an accelerating one
(Sec.~\ref{secOne}), where we find that, in the infinite
acceleration limit, the genuine tripartite entanglement saturates at
$4s^2$ ({\em i.e.}~the original entanglement encoded between the two
inertial observers).

In the scenario considered here, the fact that Leo and Nadia are
accelerated is perceived as an {\em ex novo} entanglement (function
of the acceleration) across the Rindler horizon, interlinking
indipendently both mode pair described by the corresponding
observers. The information loss is such that even if the state
contains infinite entanglement when described by inertial observers,
it contains no quantum correlations when described by two observers
traveling at finite acceleration. If one considers even higher
acceleration of the observers, it is basically the (inaccessible)
entanglement modes in the Rindler region $I$ and modes in the
Rindler region $II$ which is redistributed into genuine four-partite
form. The tripartite correlations tend to vanish as a consequence of
the thermalization which destroys the inertial bipartite
entanglement. The multipartite entanglement, obviously, increases
infinitely with acceleration because the entanglement between
causally disconnected modes increases without bound with
acceleration.  It is remarkable that such promiscuous distribution
of entanglement can occur without violating the fundamental monogamy
constraints on entanglement sharing \cite{contangle,hiroshima}.
%This
%occurs because of the `infinite' freedom that a CV system offers as
%compared for instance, to the two-dimensional Hilbert space of
%qubits \cite{unlim}}.

\begin{figure*}[t]
\centering{\includegraphics[width=13cm]{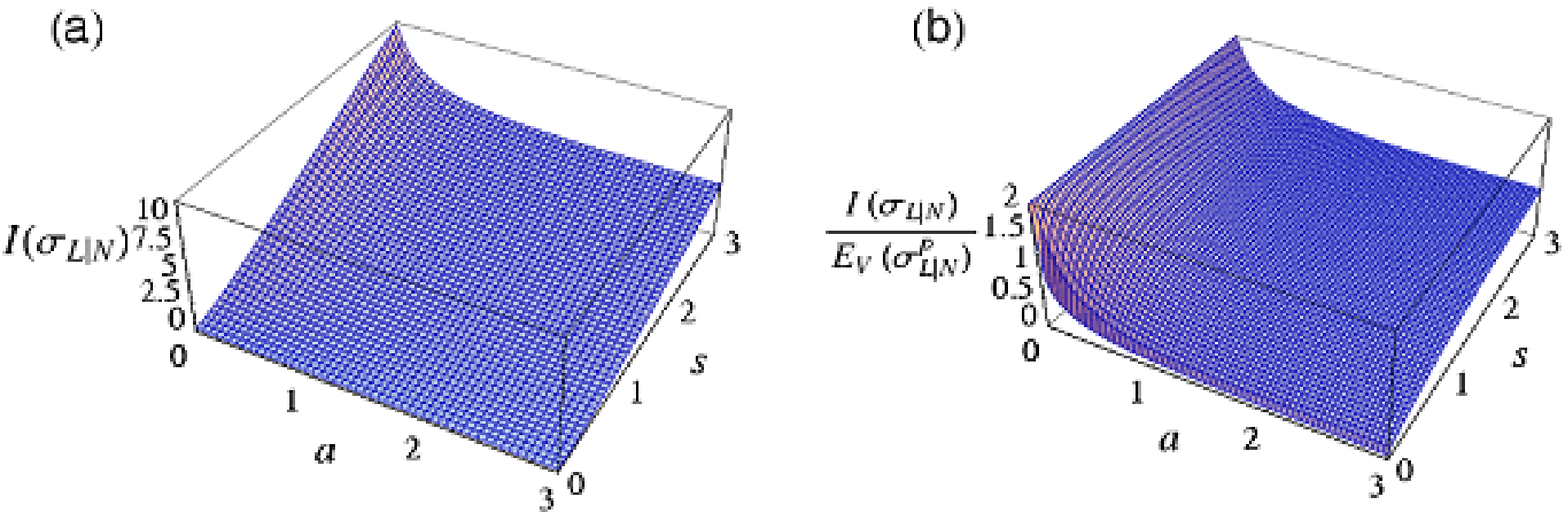}%
\caption{\label{mi4decay}(color online) Total correlations between
the modes described by the two non-inertial observers Leo and Nadia,
traveling with equal, uniform acceleration given by the effective
squeezing parameter $a$. From the inertial perspective, the field is
in a two-mode squeezed state with squeezing degree $s$. Plot (a)
shows the mutual information $I(\sig_{L|N})$, given by
Eq.~(\ref{mi4}), as a function of $a$ and $s$. In plot (b) the same
quantity is normalized to the entropy of entanglement perceived by
inertial observers, $E_V(\sig^P_{L|N})$, \eq{ev4}. Notice in (a) how
the mutual information is an increasing function of the squeezing
parameter $s$ and saturates to a nonzero value in the limit of
infinite acceleration; in contrast, the entanglement vanishes at
finite acceleration (see Fig.~\ref{fi4decay}). Plot (b), shows that
this asymptotic value is smaller than the entropy of entanglement
described by inertial observers (which is equal to the classical
correlations described by the inertial observers). Therefore,
classical correlations are also degraded when both observers are
accelerated, in contrast to the case where only one observer is in
uniform acceleration (see Fig.~\ref{fi4decay}).}}
\end{figure*}

To give a simple example, suppose the bipartite entanglement
contained in the sate described by inertial observers is given by
$4s^2 = 16$ for $s = 2$. If both observers accelerate such that the
two-mode squeezed state described from the inertial perspective has
squeezing parameter $a=7$, the four-partite entanglement (given by
\eq{tau4}), is $81.2$ ebits, more than $5$ times the inertial
bipartite entanglement. At the same time, a bipartite entanglement
of $4a^2 = 196$ is present, from a non-inertial perspective, between
modes in region $I$ and modes in region $II$. A final {\it caveat}
needs to be stated. The above results suggest that unbounded
entanglement is created by merely the observers' motion. This
requires of course an unlimited energy needed to fuel their
spaceships, let alone all the technicalities of realizing {\em in
toto} such a situation in practice. Unfortunately, this entanglement
is mostly unaccessible, as both Leo and Nadia are confined in their
respective Rindler region $I$. The only entanglement resource they
are left with is the degraded two-mode thermal squeezed state.

\subsection{Mutual information}

It is very interesting to evaluate the mutual information
$I(\sig_{L|N}$) between the states described by Leo and Nadia, both
moving with acceleration parameter $a$.

In this case the symplectic spectrum of the reduced (mixed) two-mode
CM   $\sig_{L|N}$ of \eq{sig4} is degenerate \cite{extremal},
yielding
$\eta^-_{L|N}=\eta^+_{L|N}=\left({\det\sig_{L|N}}\right)^{\frac14}$.
From \eq{migau}, the mutual information then reads
\begin{equation}\label{mi4}
I(\sig_{L|N}) = f(\sqrt{\det\sig_{L}})+f(\sqrt{\det\sig_{N}}) - 2
f\left[\left({\det\sig_{L|N}}\right)^{\frac14}\right]\,.
\end{equation}
Explicitly:\\
\begin{widetext}
\noindent $I(\sig_{L|N})= 2 \cosh ^2(a)\cosh ^2(s) \log [\cosh ^2(a)
\cosh ^2(s)]  - [\cosh (2 s) \cosh ^2(a) + \sinh ^2(a) - 1]
      \log \{\frac{1}{2} [\cosh (2 s)  \cosh ^2(a) + \sinh ^2(a) -
                  1]\}  + \frac12 \{[2 \cosh (2 s) \sinh ^2(2 a) + \cosh (4 a) + 3]^{\frac12} -
            2\} \log \{[2 \cosh (2 s) \sinh ^2(2 a) + \cosh (4 a) + 3]^{\frac12} -
            2\} - \frac12 \{[2 \cosh (2 s) \sinh ^2(2 a) + \cosh (4 a) + 3]^{\frac12}
                  +
            2\} \log \{[2 \cosh (2 s) \sinh ^2(2 a) + \cosh (4 a) + 3]^{\frac12}
            +
            2\}+\log(16)$.
\end{widetext}

We plot the mutual information both directly, and normalized to the
inertial entropy of entanglement, which is equal to \eq{ev3},
\begin{equation}\label{ev4}
E_V(\sig_{L|N}^P)= f(\cosh 2s)\,,
\end{equation}
with $f(x)$ given by \eq{entfunc}. We immediately notice another
novel effect. Not only the entanglement is completely destroyed at
finite acceleration, but also classical correlations are degraded.
(see Fig.~\ref{mi4decay}(b)). This is very different to the case
with a single non-inertial observer where classical correlations
remain invariant.

The asymptotic state described by Leo and Nadia, in the infinite
acceleration limit $a \rightarrow \infty$), contains indeed some
residual classical correlations (whose amount is an increasing
function of the squeezing $s$). But these correlations are {\em
always} smaller than the classical correlations described from the
inertial perspective, given by \eq{ev4}. Classical correlations are
robust against the effects of the double acceleration only when the
classical correlations in the state described by inertial observers
are infinite (corresponding to infinite entanglement from the point
of view of inertial observers, $s \rightarrow \infty$). The
entanglement, however, is always fragile, since we have seen that it
is completely destroyed at a finite, relatively small acceleration
parameter $a$.

\begin{figure}[t!]
\centering{\includegraphics[width=8.5cm]{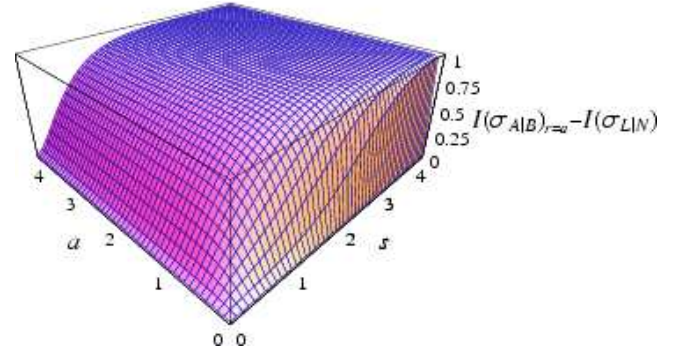}%
\caption{\label{difetto}(color online) Plot, as a function of the
acceleration parameter $a$ and the squeezing parameter $s$, of the
difference between the mutual information described by the inertial
Alice and the non-inertial Rob, and the mutual information described
by the uniformly accelerating Leo and Nadia, as given by
\eq{defect}.}}
\end{figure}
Another interesting fact is that, comparing Figs.~\ref{mi3decay}(a)
and~\ref{mi4decay}(a), one sees that in both cases (either one or
two non-inertial observers) the mutual information between the two
``real'' observers is a function of the acceleration parameter and
of the initial squeezing. In the case of both accelerated observers,
however, the mutual information is always smaller, as we have just
discussed. It is interesting to study the difference between them
(where we set for ease of comparison equal acceleration parameters,
$r=a$, where $r$ regulates Rob's acceleration when Alice is
inertial, and $a$ is related to the acceleration of both Leo and
Nadia in the present situation),
\begin{equation}\label{defect}
D(a,s) = \left.I(\sig_{A|R})\right|_{r=a} - I(\sig_{L|N})\,.
\end{equation}
The quantity $D(a,s)$ is plotted in Fig.~\ref{difetto}:
surprisingly, it is strictly bounded. It increases both with $s$ and
$a$, but in the asymptotic limit of infinite inertial entanglement,
$D(a,s \rightarrow \infty)$ saturates exactly to $1$ (as it can be
checked analytically) for any $a > 0$. We remark that both mutual
informations $I(\sig_{A|R})$ and $I(\sig_{L|N})$ diverge in this
limit: yet their difference is finite and equal to one. Clearly, the
small deficit of the mutual information seen when both observers are
accelerated, is reflected as loss of classical correlations, as
plotted in Fig.~\ref{mi4decay}(b). Mysteriously, the Unruh
thermalization affects classical correlations when both observers
are accelerated: however, it degrades at most one absolute unit of
classical correlations. This means that in the case when both Leo
and Nadia escape the fall into a black hole, not only their
entanglement is degraded but {\em there is also a loss of classical
information} \cite{letter}.

\section{Discussion and outlook}\label{SecConcl}

We presented a thorough study of classical and quantum correlations
between modes of a scalar field described by observers in uniform
acceleration. By considering the state of the field in the simplest
multi-mode squeezed state possible (the two-mode case) from the
perspective of inertial observers, we were able to investigate in
detail the entanglement in all partitions of the system from
non-inertial perspectives, specifically when one observer is in
uniform acceleration and when both of them are accelerated. We find
that in both settings the accessible entanglement is degraded with
the observers' acceleration and we explain this degradation as an
effect of re-distribution of the entanglement in the state described
from an inertial perspective.

Our main results can be summarized as follows. When one of the
observers is accelerated the entanglement lost between the modes
described by him and the inertial observer is re-distributed into
tripartite correlations. No entanglement is generated between the
mode described by the inertial observer and the modes in the
causally disconnected region $II$. This shows that indeed the
behavior for bosonic fields is very different to the Dirac case
where the entanglement lost from the perspective of non-inertial
observers is re-distributed not into tripartite correlations but
into bipartite correlations between the mode described by the
inertial observer and the mode in region $II$. The analysis of the
mutual information shows that in this case classical correlations
are conserved independently of the acceleration. The situation
changes drastically by considering that both observers are
accelerated. In this case the entanglement lost from the
non-inertial perspective is re-distributed into mainly four-partite
correlations although some tripartite correlations exist for finite
acceleration. The surprising result here (though expected in the
framework of distributed entanglement, as the additional fourth mode
comes into play) is that entanglement vanishes completely at a
finite acceleration. This is also drastically different to the
results in the Dirac case where entanglement remains positive for
all accelerations (as a direct consequence of the restricted Hilbert
space in that instance). Another surprising result is this case is
that we find that classical correlations are no longer invariant to
acceleration but are also degraded to some extent. We analyzed the
entanglement between the modes of the field described by two
accelerated observers as a function of their frequencies, and found
that for a fixed acceleration high frequency modes remain entangled
while lower frequency modes disentangle. In the limit of infinitely
accelerated observers, the field modes are in a separable state for
any pair of frequencies.

The tools developed in this paper can be used to
 investigate the problem of information loss in black
 holes \cite{letter}. There is a correspondence between the
 Rindler-Minkowski frames and the Schwarzschild-Kruskal frames \cite{unruh,birelli} that
 allows us to study the loss (and re-distribution) of quantum and
 classical correlations for observers describing entangled modes outside the black hole,
 extending and re-interpreting the results presented in
 Sec.~\ref{secTwo} of this paper. In that case the degradation of
 correlations can be understood as essentially being due to the
 Hawking effect \cite{information}.

Furthermore, all our results can be in principle corroborated
experimentally in a quantum-optics setting. The role of the
acceleration on the description of the field can be reproduced by
the effects of a nonlinear crystal through the mechanism of
parametric down-conversion \cite{walls}. The results of
Sec.~\ref{secTwo}, for instance, can be applied to study the
efficient generation and the entanglement characterization in
four-mode Gaussian states of light beams \cite{unlim}. In such a
setting each of the modes can be really accessed and manipulated and
the different types of entanglement can be described by true
observers and employed as a resource for bipartite and/or
multipartite transmission and processing of CV quantum information
\cite{adebook,brareview}.

We are currently interested in the study of classical and quantum
correlations in general multi-mode squeezed states which involve
several modes being pair-wise entangled \cite{birelli}. The study of
entanglement in this state will provide a deeper understanding of
quantum information in quantum field theory in curved spacetimes.

%\addcontentsline{toc}{section}{Acknowledgments}
\section*{Acknowledgements} \noindent
We are indebted to Mauro Paternostro for pointing Ref.~\cite{ahn} to
our attention. We thank Fabrizio Illuminati, Rob Mann, Frederic
Schuller, Orlando Luongo and Jian-Yang Zhu for very fruitful
discussions.  GA and IFS acknowledge the warm hospitality of the
Centre for Quantum Computation, Cambridge (UK), where this work was
started. ME acknowledges financial support from The Leverhulme
Trust. This work is supported by the European Union through the
Integrated Project QAP (IST-3-015848), SCALA (CT-015714), and
SECOQC.

\appendix
\section{Upper bound on the mixed-state tripartite
entanglement in presence of two accelerated observers}
\label{secApp}

We are interested in computing the residual tripartite contangle,
\eq{taures}, distributed among modes described by observers
anti-Leo, Leo and Nadia in the reduced mixed three-mode state
$\sig_{\bar L L N}$ obtained from \eq{sig4} (with $l=n\equiv a$) by
tracing over the degrees of freedom of anti-Nadia. To quantify such
tripartite entanglement exactly, it is necessary to compute the
three-mode bipartite contangle between one mode and the block of the
two other modes. This requires solving the nontrivial optimization
problem of \eq{tau} over all possible pure three-mode Gaussian
states. However, from the definition itself \eq{tau}, the bipartite
contangle $\tau (\sig_{i\vert (jk)})$ (with $i,j,k$ a permutation of
$\bar L, L, N$) is bounded from above by the corresponding bipartite
contangle $\tau (\sig_{i\vert (jk)}^p )$ in any pure, three-mode
Gaussian state with CM $\sig_{i\vert (jk)}^p \le \sig_{i\vert
(jk)}$. As an ansatz we can choose pure three-mode Gaussian states
whose CM $\sig_{\bar L L N}^p$ has the same matrix structure of our
mixed state $\sig_{\bar L L N}$ (in particular, zero correlations
between position and momentum operators, and diagonal subblocks
proportional to the identity), and restrict the optimization to such
class of states. This task is accomplished by choosing a pure state
given by the following CM \cite{unlim}
\begin{equation}\label{sigpbound}
\gr\gamma^p_{\bar L L N}=S_{\lambda_I,\lambda_{II}}(a)
S_{\lambda_I,\nu_I}(t) \id_{6} (t) S_{\lambda_I,\nu_I}^T
S_{\lambda_I,\nu_I}(t) S_{\lambda_I,\lambda_{II}}(a)\,,
\end{equation}
where we have adopted the notation of \eq{in4}, and
$$
t=\frac12{\rm arccosh}\left[\frac{1+{\rm sech}^2a\tanh ^2s}{1-{\rm
sech}^2a\tanh^2s}\right]\,. $$ We have then
\begin{equation}\label{taubnd}
\tau (\sig_{i\vert (jk)} )\le g[(m_{i\vert (jk)}^{\gamma})^2]\,,
\end{equation}
where $m^\gamma$ is meant to determine entanglement in the state
$\gr\gamma^p$, \eq{sigpbound}, via \eq{tau}. The bipartite
entanglement properties of the state $\gr\gamma^p$  can be
determined analogously to what done in Sec.~\ref{secOne}. We find
\begin{eqnarray}
 m^\gamma_{\bar N|(\bar L L)} &=& \frac{1+{\rm sech}^2a\tanh^2s}
{1-{\rm sech}^2a\tanh^2s}\,, \\
  m^\gamma_{\bar L|(LN)} &=& \cosh^2a+m^\gamma_{\bar N|(\bar L L)} \sinh^2a\,, \\
   m^\gamma_{L|(\bar LN)} &=& \sinh^2a+m^\gamma_{\bar N|(\bar L L)}
   \cosh^2a\,.
\end{eqnarray}
Eqs.~(\ref{taures},\ref{taubnd}) thus lead to
\begin{equation}\label{tribnd}
\begin{split}
\tau(\sig_{\bar L | L | N}) \le \min \big\{&g[(m_{\bar L\vert
(LN)}^{\gamma} )^2]-g[m_{\bar L\vert L}^2 ], \\ &g[(m_{N\vert (\bar
L L)}^{\gamma})^2]-g[m_{L\vert N}^2 ]\big\}\,,
\end{split}
\end{equation}
where the two-mode entanglements $m$ without the superscript
``$\gamma$'' are referred to the reductions of the mixed state
$\sig_{\bar L | L | N}$ and are listed in
Eqs.~(\ref{m411sep}--\ref{m4ln}). In \eq{tribnd} the quantity
$g[(m_{L\vert (\bar L N)}^{\gamma} )^2]-g[m_{\bar L\vert L}^2
]-g[m_{L\vert N}^2 ]$ is not included in the minimization, being
always larger than the other two terms. Numerical investigations in
the space of all pure three-mode Gaussian states seem to confirm
that the upper bound of \eq{tribnd} is actually tight (meaning that
the three-mode contangle is globally minimized on the state
$\gr\gamma^p$), but this statement can be left as a conjecture since
it is not required for the subsequent analysis of
Sec.~\ref{secTwoMulti}.

%\addcontentsline{toc}{section}{References}
%\section*{References}


\begin{thebibliography}{99}
\bibitem{noteentro} The entanglement measures defined before, like the negativities and the contangle,
are monotonically increasing functions of the reduced Von Neumann
entropy for pure Gaussian states, yielding an equivalent
quantification of bipartite pure-state entanglement.
\bibitem{notemodefreq}
Throughout the paper, the same Greek letters will be used
indifferently to denote the mode frequencies, or to label the modes
themselves.
\bibitem{notem} We refer to the notation of \eq{tau} and write, for each
partition $i|j$, the corresponding parameter $m_{i|j}$ involved in
the optimization problem which defines the contangle for bipartite
Gaussian states.
\bibitem{noteid} When writing a transformation acting on two modes, we are implicitly
assuming that the identity operation is acting on each of the two
remaining modes; here, these identity matrices will be omitted.
\end{thebibliography}
\end{document}